\documentclass[twocolumn]{aastex631}
\shorttitle{search for massive galaxy population in a protocluser at $z=2.39$}
\shortauthors{Yonekura et al.}
\graphicspath{{./}{figures/}}
\begin{document}

\title{A Search for Massive Galaxy Population in a Protocluster of LAEs at $z=2.39$ near the Radio Galaxy 53W002}

\correspondingauthor{Naoki Yonekura}
\email{yonekura@cosmos.phys.sci.ehime-u.ac.jp}

%\author[0000-0002-0786-7307]{Greg J. Schwarz}
\author{Naoki Yonekura}
\affiliation{Graduate School of Science and Engineering, Ehime University, 2-5 Bunkyo-cho, Matsuyama, Ehime 790-8577, Japan}

\author{Masaru Kajisawa}
\affiliation{Graduate School of Science and Engineering, Ehime University, 2-5 Bunkyo-cho, Matsuyama, Ehime 790-8577, Japan}
\affiliation{Research Center for Space and Cosmic Evolution, Ehime University, 2-5 Bunkyo-cho, Matsuyama, Ehime 790-8577, Japan}

\author{Erika Hamaguchi}
\affiliation{Graduate School of Science and Engineering, Ehime University, 2-5 Bunkyo-cho, Matsuyama, Ehime 790-8577, Japan}

\author{Ken Mawatari}
\affiliation{National Astronomical Observatory of Japan, 2-21-1 Osawa, Mitaka, Tokyo 181-8588, Japan}

\author{Toru Yamada}
\affiliation{Institute of Space and Astronautical Science, Japan Aerospace Exploration Agency, 3-1-1, Yoshinodai, Chuo-ku, Sagamihara, Kanagawa 252-5210, Japan}
\begin{abstract}
  We searched for massive galaxy population in the known large-scale high-density structure of Lyman$~\alpha$ emitters (LAEs) at $z=2.39$ near the radio galaxy 53W002 by using $B,~V,~i^\prime,~J,~H,~{\rm{and}}~K_s$-bands imaging data taken with Suprime-Cam and MOIRCS on the Subaru telescope.
  We selected 62 protocluster member candidates by their $JHK_s$-band colors and Spectral Energy Distribution (SED) fitting analysis ($JHK_s$-selected galaxies) in our survey field of $70.2~{\rm{arcmin}}^{2}$, and compared their physical properties estimated from the SED fitting with a comparison sample in the COSMOS field.
  We found significant number density excesses for the $JHK_s$-selected galaxies in the 53W002 field at $K_s<22.25,~J-K_s>2,~{\rm{or}}~V-K_s>4$.
  In particular the number density of the $JHK_s$-selected galaxies with $K_s<22.25~{\rm{and}}~J-K_s>2$ in the 53W002 field is nine times higher than the comparison sample.
  Most of those with $K_s<22.25~{\rm{and}}~J-K_s>2$ are massive galaxies with $M_s>10^{11}~M_\odot$, and their sSFRs of $10^{-11}$--$10^{-10}~\rm{yr^{-1}}$ suggest that the star formation has not yet stopped completely.
  We also found a density excess of quiescent galaxies with $M_s=5\times10^{10}$--$10^{11}~M_\odot~{\rm{and}}~{\rm{sSFR}}<10^{-11}~\rm{yr^{-1}}$ as well as that of low-mass galaxies with $M_s=10^{9.75}$--$10^{10}~M_\odot$ and various sSFRs.
  The massive galaxies with $M_s>10^{11}~M_\odot$ are not located at the density peaks of LAEs, but widely distributed along the similar direction with the structure of LAEs over $\sim15$--$20$ comoving Mpc.
  On the other hand, the quiescent galaxies with ${\rm{sSFR}}<10^{-11}~\rm{yr^{-1}}$ clearly avoid the structure of LAEs.
  Our results suggest that massive galaxies also exist in this protocluster discovered by the moderate overdensity of LAEs and their star formation activities depend on location in the protocluster.
\end{abstract}




%%%%%%%%%%%%%%%%%%%%%%%%%%%%%%%%%%%%%%%%%%%%%%%%%%%%%%%%%%%%%%%%%%
%%%%%%%                                                    %%%%%%%
%%%%%%%                                                    %%%%%%%
%%%%%%%                    Introduction                    %%%%%%%
%%%%%%%                                                    %%%%%%%
%%%%%%%                                                    %%%%%%%
%%%%%%%%%%%%%%%%%%%%%%%%%%%%%%%%%%%%%%%%%%%%%%%%%%%%%%%%%%%%%%%%%%
\section{Introduction} \label{sec:intro}
In the present universe, galaxy clusters are dominated by massive early-type galaxies, while there are many spiral galaxies in field environments \citep[e.g.,][]{Dressler_1980,Goto_2003}.
Those early-type galaxies in clusters show similarly red optical colors and form the color-magnitude relation or red sequence in the color-magnitude plain \citep[e.g.,][]{Bower_1992}.
Such red sequences of massive cluster members have been discovered at progressively higher redshifts up to $z\sim2$ \citep[e.g.,][]{Aragon_1993,Stanford_1998,Snyder_2012,Willis_2020}.
The tightness of their red colors suggests that stars in these massive cluster galaxies were formed at $z>2$ \citep{Bower_1992}.
In the $\Lambda$CDM paradigm of structure formation, it is predicted that such clusters have been formed hierarchically from smaller halos, and their progenitors, namely protoclusters, consist of hundreds of halos and extend over tens of comoving Mpc at $z>2$ \citep[e.g.,][]{Chiang_2013,Muldrew_2015}.
Therefore, direct observations of protoclusters at $z>2$ are important to understand the formation and evolution of massive cluster galaxies. 

Targeting high-redshift radio galaxies is an efficient way to find protoclusters.
These galaxies tend to be the most massive galaxies with $M_s>10^{11}\ M_\odot$ in the early universe \citep[e.g.,][]{Seymour_2007}, which suggests that they can be good markers of high-density peaks, and actually reside in high-density environments more frequently than normal (non-AGN) similarly massive galaxies \citep{Hatch_2014}.
Many narrow-band imaging observations searching for line emitters around radio galaxies have successfully discovered protoclusters at $z>2$ \citep[e.g.,][]{LeFevre_1996,Pascarelle_1996b,Kurk_2000,Pentericci_2000,Venemans_2002,Venemans_2004,Venemans_2005,Venemans_2007,Kuiper_2011,Hayashi_2012,Cooke_2014,Husband_2016}.
Overdensities of Lyman break galaxies (LBGs) around the radio galaxies have been found by the broad-band color selection technique \citep[e.g.,][]{Miley_2004,Overzier_2006,Overzier_2008}.
The similar near-infrared (NIR) color-selection methods for Balmer break galaxies were used to search high-density regions of massive galaxies in the radio galaxy fields \citep[][]{Kajisawa_2006,Hatch_2011a}.

Wide-field surveys with the narrow-band imaging and/or the broad-band color selection technique mentioned above have also discovered high-redshift protoclusters which are not biased to the radio galaxy fields \citep[e.g.,][]{Steidel_1998,Steidel_2000,Steidel_2005,Shimasaku_2003,Ouchi_2005,Geach_2012,Toshikawa_2012,Toshikawa_2016,Toshikawa_2018,Lee_2014,Lemaux_2014,Diener_2015,Badescu_2017,Jiang_2018,Higuchi_2019}.
The Lyman break technique with wide-field optical imaging can search a large volume effectively, but the spectroscopic confirmation is essential to avoid the chance alignment.
The line emitter searches with narrow-band imaging have the advantage of selecting galaxies with a narrow redshift window (e.g., $\Delta z\sim0.05$), while their survey volumes tend to be limited.
The wide-field optical narrow-band observations searching for Lyman $\alpha$ emitters (LAEs) can both mitigate the limited survey volume and keep effectiveness of picking up high-density regions of galaxies at the same redshifts.

On the other hand, LAEs typically have a low stellar mass of $\sim10^{9}\ M_\odot$ and young stellar age of several tens of Myr \citep[e.g.,][]{Gawiser_2007}.
LBGs and other line emitters such as H$\alpha$ and [OIII] are basically star-forming galaxies.
Therefore, these selection methods could miss massive galaxies and/or quiescent ones with little star formation, which are similar with massive early-type galaxies in the present clusters.
Since rest-frame optical--NIR SEDs give us the information of old stellar population of galaxies, the NIR follow-up observations for protoclusters are important to understand the formation and growth of massive galaxies in the high-density environments.
In particular, such follow-up observations are essential to confirm whether massive galaxies have already formed or not in the high-density structures of LAEs, which tends to be very young and less dusty low-mass galaxies.

However, the number of such NIR follow-up studies for protoclusters is still limited.
\citet{Uchimoto_2012} and \citet{Kubo_2013} found that NIR-selected massive galaxies show a significant number density excess along a large-scale structure of LAEs in the SSA22 field, which is one of the largest protoclusters discovered so far \citep{Yamada_2012}.
\citet{Kubo_2015,Kubo_2021} spectroscopically confirmed that many of the massive galaxies are protocluster members.
\citet{Shi_2019a} also found a significant excess of NIR-selected Balmer break galaxies in a rich structure of LAEs and LBGs at $z=3.78$ \citep{Lee_2014,Dey_2016}.
Furthermore, several studies reported a segregation in the spatial distribution between LAEs and more massive galaxies in protoclusters \citep[e.g.,][]{Shimakawa_2017b,Shi_2019a,Shi_2020}.
It is still unclear how massive galaxies (if they exist) are distributed in other protoclusters traced by LAEs.

In this study, we use optical and NIR imaging data taken with Suprime-Cam and MOIRCS on the Subaru telescope to search for massive galaxies in a large-scale high-density structure of LAEs at $z=2.39$ discovered by \citet{Mawatari_2012} near the radio galaxy 53W002.
The 53W002 protocluster was initially discovered as an overdensity of 18 LAEs selected by the intermediate-band excess of HST/WFPC2 data around the radio galaxy \citep{Pascarelle_1996a,Pascarelle_1996b}.
\citet{Keel_1999} carried out a wider ground-based imaging observation over $14\farcm3\times14\farcm3$ area with a similar intermediate-band filter and found 14 LAE candidates.
\citet{Yamada_2001} performed a NIR imaging observation in the WFPC2 field.
They found that the LAEs are very faint or not detected in the NIR data and there are few massive quiescent galaxy candidates at the redshift with bright NIR magnitudes and red colors.
\citet{Mawatari_2012} observed $31^{\prime}\times24^\prime$ field with Suprime-Cam and a custom-made narrow-band filter for the Ly$\alpha$ emission at $z\sim2.39$.
They discovered the large-scale high-density structure of LAEs north or northeast of 53W002, and found that the radio galaxy is located at outskirt of the structure.
Number density of LAEs in the highest-density region of $5^{\prime}\times4^{\prime}$ is four times higher than the average value over the Suprime-Cam field.
This overdensity is not as extremely high as the SSA22 protocluster, and corresponds to a moderately rich structure.
We search for massive galaxies over a $\sim70$ arcmin$^{2}$ field along the structure of LAEs with the optical--NIR imaging data, and compare their statistical properties with similarly selected field galaxies in the COSMOS field.

\begin{figure}[th!]
  \includegraphics[width=8.5cm]{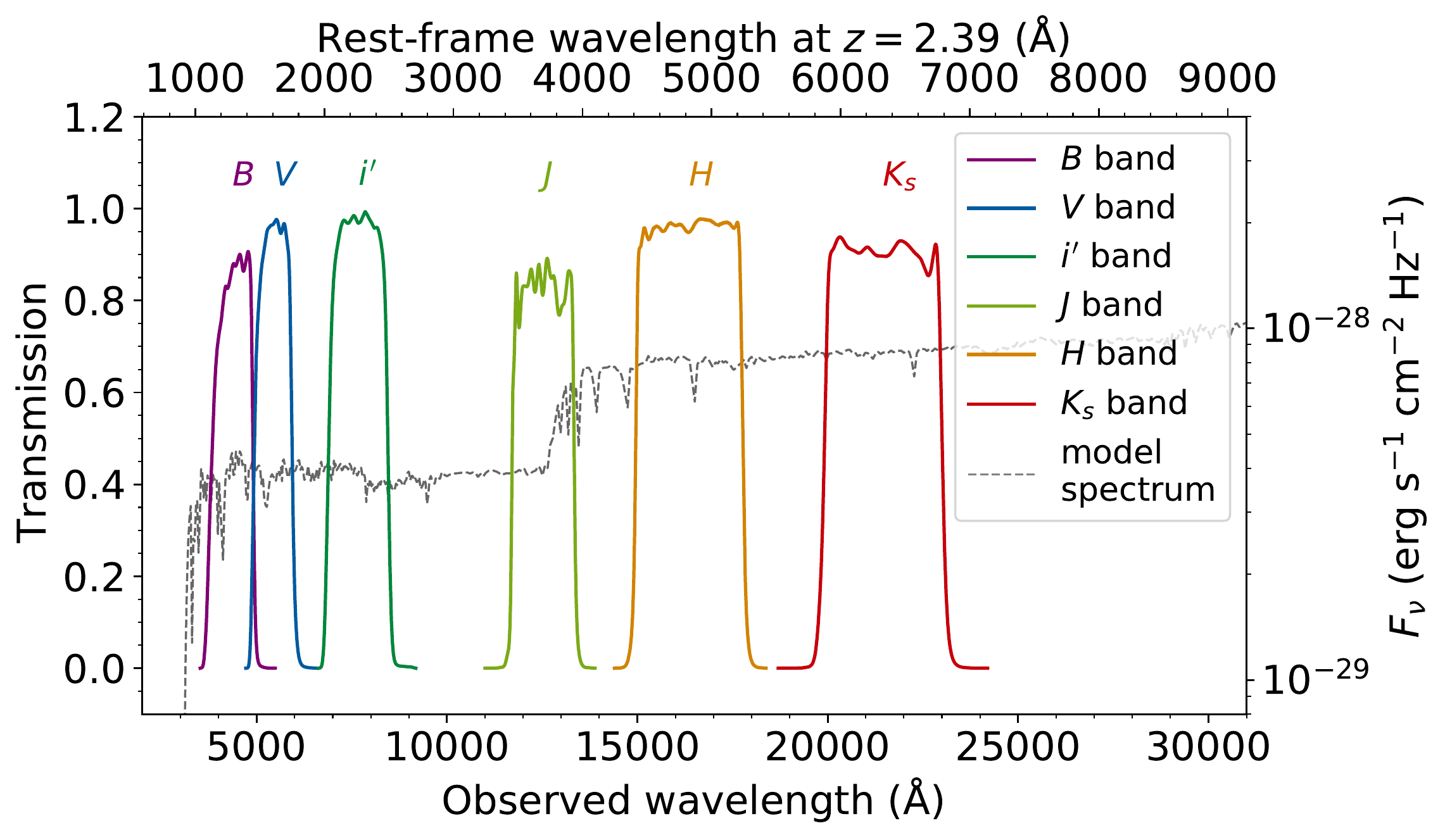}
  \caption{Transmission curves of filters used in this study and a model spectrum at $z=2.39$. The color solid lines show the transmission curves of the filters. Grey dashed line shows the model spectrum at $z=2.39$ from \citet{Bruzual_2003} with age of 1 Gyr and star-formation history of SFR $\propto\exp(-t/{\rm{1~Gyr}})$.}
  \label{fig:fig1}
\end{figure}

This paper is organized as follows: We describe observations and data reduction in Section \ref{sec:data}.
The procedures of selecting candidates for protocluster members at $z=2.39$ and estimating their physical properties are described in Section \ref{sec:analysis}.
We present results in Section \ref{sec:results} and discuss them in Section \ref{sec:discussion}.
The summary is given in Section \ref{sec:summary}.
We use the AB magnitude system \citep{Oke_1983}.
%except for the NIR color selection criteria defined by Vega magnitude in our previous study.
%Conversions between the Vega and AB system in the $J,H,$ and $K_s$ bands of MOIRCS are 
%$J_{\rm{AB}} = J_{\rm{Vega}}+0.915$, 
%$H_{\rm{AB}} = H_{\rm{Vega}}+1.354$, and
%$K_{s\ \rm{AB}} = K_{s\ \rm{Vega}}+1.834$, respectively.
We assume cosmological parameters of $H_0=70\ {\rm{km}\ {\rm{s^{-1}}}\ {Mpc^{-1}}}$, $\Omega_{\rm{m}}=0.3$, and $\Omega_{\rm{\Lambda}}=0.7$.

%%%%%%%%%%%%%%%%%%%%%%%%%%%%%%%%%%%%%%%%%%%%%%%%%%%%%%%%%%%%%%%%%%
%%%%%%%                                                    %%%%%%%
%%%%%%%                                                    %%%%%%%
%%%%%%%                      Data                          %%%%%%%
%%%%%%%                                                    %%%%%%%
%%%%%%%                                                    %%%%%%%
%%%%%%%%%%%%%%%%%%%%%%%%%%%%%%%%%%%%%%%%%%%%%%%%%%%%%%%%%%%%%%%%%%

\section{Observations \& Data Reduction} \label{sec:data}

\subsection{Data and Reduction}
In this study, we used the $B$, $V$, $i^{\prime}$, $J$, $H$, and $K_s$-band imaging data.
Figure \ref{fig:fig1} shows the transmission curves of the used filters and a model spectrum of a galaxy at $z=2.39$.
The wavelengths of $B$, $V$, and $i^{\prime}$ bands correspond to the ultraviolet in the rest frame at $z=2.39$.
Those of $J$, $H$, and $K_s$ bands correspond to the rest-frame optical at $z=2.39$, and the Balmer/4000$\rm{\AA}$ break falls between the $J$ and $H$ bands at the redshift. 
The summary of observations is given in Table \ref{tab:table1}.

\begin{figure}[ht!]
  \includegraphics[width=8.5cm]{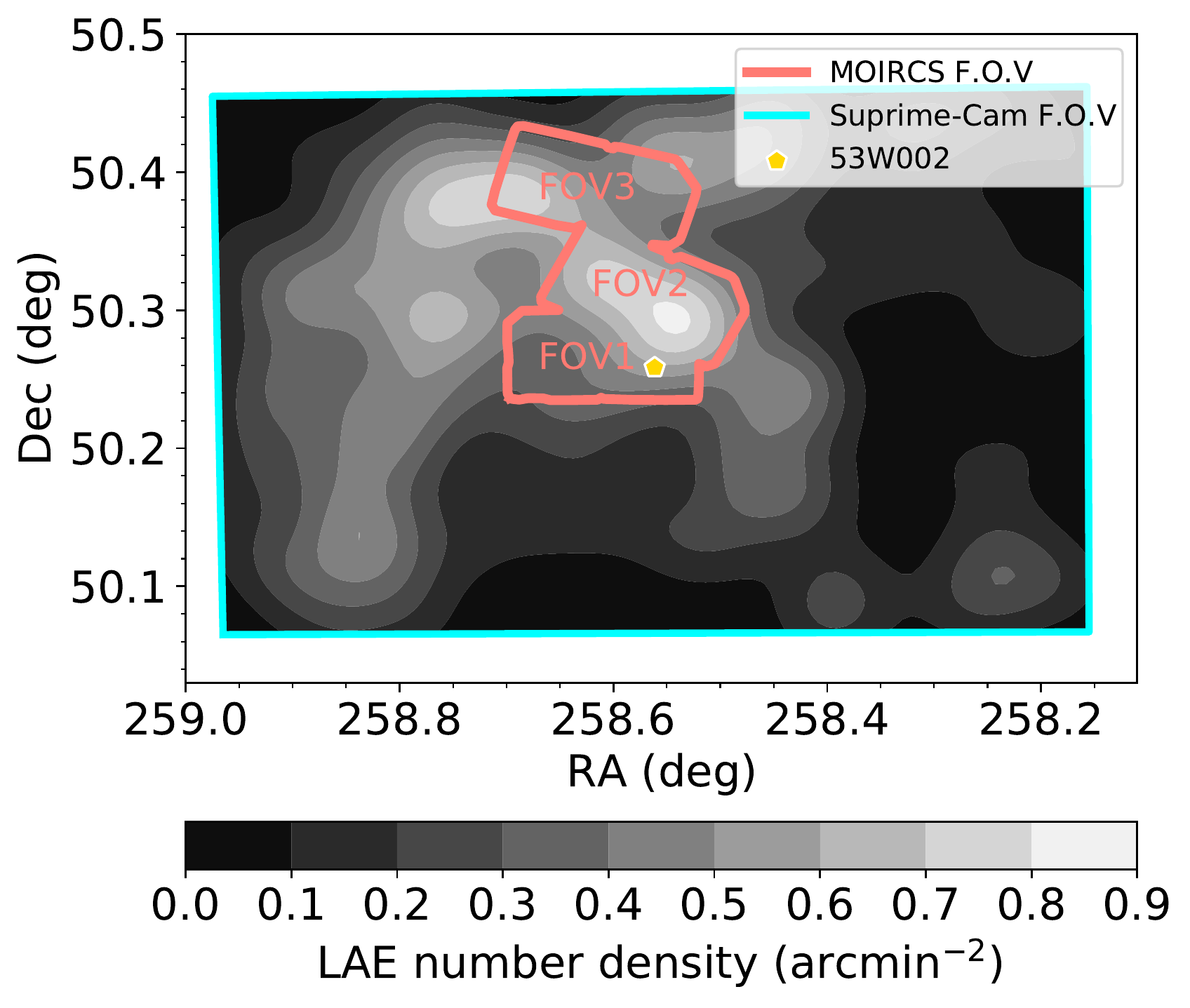}
  \caption{Fields of view of Suprime-Cam and MOIRCS observations. 
  The blue solid line shows the Suprime-Cam field of view, while the red solid line shows the MOIRCS field of view. The grey contour shows the LAE number density smoothed with a Gaussian kernel with $\sigma=1\farcm5$ from \citet{Mawatari_2012}.}
  \label{fig:fig2}
\end{figure}

The $B,V,$ and $i^{\prime}$-band images were observed by Subaru Prime Focus Camera \citep[Suprime-Cam:][]{Miyazaki_2002} mounted on the 8.2 m Subaru telescope.
Figure \ref{fig:fig2} shows the fields of view of the observations and the spatial distribution and number density map of LAEs discovered by \citet{Mawatari_2012}.
The blue solid line indicates the field of view of the optical observations, and they cover the large-scale structure of LAEs.
The $B$-band image was provided by \citet{Mawatari_2012}.
The $V$ and $i^\prime$-band observations were done on 2015 April 15 and 2016 May 16 with the same field of view (centered at RA=17:14:15.2, Dec=+50:15:15.2) as the $B$-band image.
The weather condition on 2015 April 15 was cloudy, and only $i^\prime$-band imaging with 30 min on-source exposure was done under $0\farcs80$-$1\farcs4$ seeing.
On 2016 May 16, we took both $V$ and $i^\prime$-band data through thin clouds under $0\farcs6$--$1\farcs2$ seeing.
We reduced the $V$ and $i^{\prime}$-band images by using SDFRED version 2.0 \citep[][]{Yagi_2002,Ouchi_2004}.
Several frames with low S/N due to the bad condition were excluded from the final stacking of the images. 
The total exposure time for the combined $V$ and $i^{\prime}$-band images are 1.3 and 1.5 hr, respectively.
The PSF FWHMs of the reduced $V$ and $i^\prime$-band images are $0\farcs87$ and $1\farcs1$, respectively.
These observations were done in the non-photometric conditions, and therefore we calibrated these data by comparing optical and NIR colors of stars in the observed data with those of Landolt standard stars \citep{Landolt_1992}.

The $J$, $H$, and $K_s$-band data were taken with Multi Objects InfraRed Camera and Spectrograph \citep[MOIRCS:][]{Ichikawa_2006,Suzuki_2008} on the Subaru telescope.
MOIRCS covers a field of view of $4^{\prime}\times7^{\prime}$ by two HgCdTe HAWAII-2 detectors, namely chip-1 and chip-2, with a pixel scale of $0\farcs117$ pixel$^{-1}$.
As shown in Figure \ref{fig:fig2}, we observed three fields of view of MOIRCS, namely, FOV1, FOV2, and FOV3.
The FOV1 is the most southern one and contains the radio galaxy 53W002.
The FOV2 covers 53W002F-HDR, a high-density region of LAEs at $z=2.39$, discovered by \citet{Mawatari_2012}.
The FOV3 is located north of FOV2 and covers another high-density region of LAEs northeast of 53W002F-HDR.
The FOV1 was observed on 2006 July 22-23 and 2007 April 27, while the FOV2 and FOV3 were observed on 2013 April 23-24 and 2014 June 13.
The observations of the FOV1 were carried out in photometric and good-seeing conditions and one of UKIRT Faint Standard stars, FS27 \citep{Leggett_2006} was also observed in the same nights for calibration.
The other two fields were observed in non-photometric conditions under $0\farcs5$-$0\farcs7$ seeing and no standard star was available for these data.

We reduced the $J, H,$ and $K_s$-bands data with MCSRED \citep{Tanaka_2011}.
The reduction methods were basically the same as those described in \citet{Kajisawa_2011}. 
In addition to basic procedures such as flat-fielding, median-sky subtraction, distortion correction, shift, and adding, the defringing process described in \citet{Kajisawa_2011} was applied for the FOV1 data.
The defringing process was not applied for the FOV2 and FOV3 data because ``fringe-free'' filters with wedged substrates were used in the observations of these two fields. 
Since no standard star was available for the FOV2 and FOV3 data, we estimated zero-point magnitudes for these data from those of the FOV1 data by matching fluxes of objects in the overlapping regions among the different fields. 
Then we additionally corrected zero-point offsets among the different bands in the same field by using the color tracks of Galactic stars.

Since we select our sample by $JHK_s$-band colors and then apply an additional criterion based on the optical-NIR SED fitting, we performed PSF matching to measure NIR and/or optical-NIR colors.
The PSF FWHMs of the $V$ and $i^\prime$-bands data are larger than those of the NIR data (Table \ref{tab:table1}), and matching the PSFs of the NIR data to the optical one leads to unnecessary loss of S/N ratio in the NIR color selection described in the next section.
In order to avoid this problem, we made two kinds of images: 
one is the images whose PSFs are matched among only the NIR data, and the other is those whose PSFs are matched to the worst one ($i^{\prime}$ band).
First, we matched PSF among the $J$, $H$, and $K_s$-band images to measure the NIR colors for the color selection.
By using {\tt{IRAF/PSFMATCH}}, we matched the PSF FWHMs of all the NIR images to $0\farcs77$, which corresponds to that of the $H$-band image in the FOV2 chip-1 smoothed with a Gaussian kernel with $\sigma=1.5$ pixel.
Second, we matched PSF FWHMs of the $B$, $V$, $i^{\prime}$, $J$, $H$, and $K_s$-band images for the SED fitting.
We aligned the $B$, $V$, and $i^{\prime}$-band images with the $K_s$-band image frame by using the {\tt{IRAF/GEOTRAN}}.
After that, we matched the PSFs of all the images to that with a FWHM of $1\farcs2$, which corresponds to that of the $i^\prime$-band image smoothed with a Gaussian kernel.

The limiting magnitudes of the final combined and PSF-matched data for each field and chip are summarized in Table \ref{tab:table2}.

\begin{deluxetable*}{ccccccc}
  %\tablenum{1}
  \tablecaption{Summary of observed data used in this study.\label{tab:table1}}
  \tablewidth{0pt}
  \tablehead{
  \colhead{Band}& \colhead{Instrument} & \colhead{Field} & \colhead{Date} &
  \colhead{\begin{tabular}{c}Exposure-\\time\\(hr)\end{tabular}} & 
  \colhead{\begin{tabular}{c}PSF FWHM\tablenotemark{a}\\($^{\prime\prime}$)\end{tabular}} &
  \colhead{Reference} 
  }
  %\decimalcolnumbers
  \startdata
  $B$ & Suprime-Cam/Subaru & & 2009 May 22-23 & 1.0 & $0.86$ & \citet{Mawatari_2012}\\\smallskip
  $V$ & Suprime-Cam/Subaru & & 2016 May 16 & 1.3 & $0.87$   & \\\smallskip
  $i^{\prime}$ & Suprime-Cam/Subaru & & 2015 April 15, 2016 May 16 & 1.5 & $1.1$  & \\
  $J$ & MOIRCS/Subaru & FOV1 & 2006 July 22-23, 2007 April 27& 0.6 & $0.39/0.38$ & \\
   &  & FOV2 & 2013 April 24-25, 2014 June 13& 1.1 & $0.61/0.63$ & \\\smallskip
   &  & FOV3 & 2013 April 23-24, 2014 June 13& 1.2 & $0.59/0.59$ & \\
  $H$ & MOIRCS/Subaru & FOV1 & 2006 July 22-23, 2007 April 27& 0.7 & $0.45/0.42$ & \\
  &  & FOV2 & 2013 April 24-25, 2014 June 13& 1.1 & $0.63/0.59$ & \\\smallskip
  &  & FOV3 & 2013 April 23-24, 2014 June 13 & 1.0 & $0.49/0.49$ & \\
  $K_s$ & MOIRCS/Subaru & FOV1 & 2006 July 22-23, 2007 April 27& 0.7 & $0.49/0.42$ &\\
  &  & FOV2 & 2013 April 24-25, 2014 June 13& 1.3 & $0.48/0.52$ &\\
  &  & FOV3 & 2013 April 23-24, 2014 June 13  & 0.8 & $0.48/0.47$ &\\
  \enddata
  \tablenotetext{a}{In the NIR data, PSF FWHM values for the two detectors of MOIRCS, namely, chip-1 (left part of the field of view) and chip-2 (right part) in each field are shown.}
\end{deluxetable*}
\begin{deluxetable*}{r c ccc c cccccc}
  %\tablenum{1}
  \tablecaption{Limiting magnitudes in each field.\label{tab:table2}}
  \tablewidth{0pt}
  \tablehead{
    \colhead{} & & 
    \multicolumn{3}{c}{$1\farcs6$ aperture \tablenotemark{a}} & & 
    \multicolumn{6}{c}{$2\farcs0$ aperture \tablenotemark{b}}
    \\ \cline{3-5} \cline{7-12}
    \colhead{Field} & &
    \colhead{$J$} & \colhead{$H$} & \colhead{$K_s$}& &
    \colhead{$B$} & \colhead{$V$}  & \colhead{$i^{\prime}$}& 
    \colhead{$J$} & \colhead{$H$} & \colhead{$K_s$}
  }
  %\decimalcolnumbers
  \startdata
  FOV1 chip-1 & &
  23.9 & 23.6 & 23.5 & & 
  28.3 & 27.6 & 26.9 & 24.9 & 24.6 & 24.5 \\\smallskip
  \textcolor{white}{FOV1} chip-2 & &
  23.8 & 23.4 & 23.4 & &
  28.3 & 27.7 & 26.9 & 24.8 & 24.4 & 24.6 \\
  FOV2 chip-1 & &
  24.0 & 23.5 & 23.7 & & 
  28.3 & 27.6 & 26.9 & 25.0 & 24.4 & 24.6 \\\smallskip
  \textcolor{white}{FOV2} chip-2 & &
  23.8 & 23.3 & 23.4 & &
  28.3 & 27.6 & 26.9 & 24.7 & 24.2 & 24.4 \\
  FOV3 chip-1 & &
  23.9 & 23.4 & 23.3 & & 
  28.4 & 27.6 & 26.9 & 24.9 & 24.3 & 24.3 \\\smallskip
  \textcolor{white}{FOV3} chip-2 & &
  23.6 & 23.3 & 23.3 & &
  28.4 & 27.7 & 27.0 & 24.6 & 24.2 & 24.2
  \enddata
  \tablenotetext{a}{5$\sigma$ limiting magnitude.}
  \tablenotetext{b}{2$\sigma$ limiting magnitude.}
  %\tablecomments{}
\end{deluxetable*}

\vspace{-17mm}
\subsection{Source Detection and Photometry}\label{subsec:detection}
%%%%%%%% detection
We carried out the source detection in the $K_s$-band images by using the software SExtractor version 2.5.0 \citep{Bertin_1996}.
A detection threshold of 1.2-times the local background root mean square over 15 connected pixels was used.
We used a region where the exposure time was more than half of maximum value in the $JHK_s$-band images to keep homogenous depth in each field, and the exposure maps were used for WIGHT\_MAP of SExtractor.
The Kron magnitude from SExtractor was used as the total $K_s$-band magnitude of detected objects.
We used objects with $K_s<23.3$, which is the brightest 5$\sigma$ limiting magnitude in the MOIRCS fields.
Finally, we detected 2381 objects in the area of 70.2 arcmin$^{2}$. 
%%%%% photometry
In order to measure colors, we performed aperture photometry with $1\farcs6$ and $2\farcs0$ diameter apertures on the images with PSF matched among the $JHK_s$ and $BVi^{\prime}JHK_s$ bands by using dual image mode of SExtractor, respectively.

In order to estimate the background noise, we made sky images by masking object pixels in the data with pseudo noise images, which was made from the sky region of the original images.
We then measured fluxes at many random positions in the sky images with the same $1\farcs6$ and $2\farcs0$ diameter apertures as the object photometry, and used their standard deviation as the background error.
In Section \ref{subsec:completeness}, we estimate the detection completeness for our sample galaxies selected with the color criteria and SED fitting.

%%%%%%%%%%%%%%%%%%%%%%%%%%%%%%%%%%%%%%%%%%%%%%%%%%%%%%%%%%%%%%%%%%
%%%%%%%                                                    %%%%%%%
%%%%%%%                                                    %%%%%%%
%%%%%%%                      Analysis                      %%%%%%%
%%%%%%%                                                    %%%%%%%
%%%%%%%                                                    %%%%%%%
%%%%%%%%%%%%%%%%%%%%%%%%%%%%%%%%%%%%%%%%%%%%%%%%%%%%%%%%%%%%%%%%%%
\section{Analysis} \label{sec:analysis}
\subsection{Sample Selection}\label{subsec:sample}
We basically selected member galaxy candidates in the 53W002 protocluster at $z\sim2.39$ by a color-selection method with $JHK_s$ bands proposed by \citet{Kajisawa_2006}.
This color selection is defined by the Vega magnitude system, and can be expressed in the AB magnitude system\footnote{Conversions between the Vega and AB magnitude system in the $J,H,$ and $K_s$ bands of MOIRCS are $J_{\rm{AB}} = J_{\rm{Vega}}+0.915$, $H_{\rm{AB}} = H_{\rm{Vega}}+1.354$, and $K_{s\ \rm{AB}} = K_{s\ \rm{Vega}}+1.834$, respectively.} as follows:
\begin{eqnarray}
  \begin{array}{c}
   (J-K_s)_{\rm{AB}} > 0.581
   \\
   \rm{and}
   \\
   (J-K_s)_{\rm{AB}} > 2(H-K_s)_{\rm{AB}}+0.541 .
   \label{eq:eq1}
  \end{array}
\end{eqnarray} 
This selection is based on the fact that the Balmer/4000$\rm{\AA}$ break falls between the $J$ and $H$ bands at $2\lesssim z \lesssim 3$.
With this selection, we can pick up not only passively evolving galaxies but also star-forming galaxies at $z\sim2.5$ except for very young ones without significant Balmer/4000$\rm{\AA}$ break.
The effect of dust extinction on this selection is small because the reddening vector is parallel to these criteria \citep[Figure 2 in][]{Kajisawa_2006}.
In this selection, we removed objects with S/N$<$3 in $H$ band, because these objects have only lower limit of the $H-K_s$ color and we cannot ensure that they satisfy the second criterion in equation (\ref{eq:eq1}).
By using this color selection, we selected 80 objects with $K_s<23.3$ in the 53W002 field.
The radio galaxy 53W002 does not satisfy the $JHK_s$-color criteria due to the effect of very strong emission lines from AGN \citep{Motohara_2001}.
\citet{Kajisawa_2011} performed this color selection in the MOIRCS Deep Survey and confirmed that selected objects have a redshift distribution with a strong peak at $z\sim2.5$ and relatively small contamination from low redshifts.
However, our observations are shallower than that, so it is possible that foreground/background objects come into the sample due to the photometric error.
In order to reduce the contamination, we set an additional criterion by using the SED fitting with the optical and NIR photometry.

%%%% SED fitting
For the SED fitting, we used $2\farcs0$ aperture photometry in the $B$, $V$, $i^{\prime}$, $J$, $H$, and $K_s$ bands.
As template spectra, we used the stellar population synthesis library of the GALAXEV \citep{Bruzual_2003}.
We assumed the initial mass function of \citet{Chabrier_2003} and three kinds of metallicities ($Z$=0.004, 0.008, and 0.02).
The assumed star-formation histories were exponentially declining SFR (SFR $\propto e^{-t/\tau}$, hereafter $\tau$ model), delayed-$\tau$ exponentially declining SFR (SFR $\propto t\times e^{-t/\tau}$, hereafter delayed $\tau$ model), single burst, and constant SFR.
The timescale $\tau$ for the $\tau$ and delayed $\tau$ models ranges from 0.001 to 10 Gyr.
The duration of star formation in the single burst model ranges from 0.01 to 1.8 Gyr.
The range of age is between 50 Myr and the cosmic age at each redshift.
The redshift range is between 0.01 and 4.99.
We used the extinction law of \citet{Calzetti_2000} with a range of $E(B-V)=0.0$--$2.0$.
We also assumed the IGM absorption of \citet{Madau_1995}.
We note that the nebular emission is not included in the model templates. 
Strong nebular emission lines such as [OII]$\lambda3727$, [OIII]$\lambda\lambda4959,5007$, H$\alpha$, and so on could significantly affect the SED of galaxies, in particular, for those with a very young age.
With the COSMOS2015 catalog described in the next subsection, we found that including the nebular emission in the templates leads to catastrophic failure in the redshift estimation with our 6-band photometric SEDs for some (relatively small) fraction of sample galaxies, while excluding the nebular emission rarely causes such failure.
Therefore, we chose not to include the nebular emission in the model templates and checked the effects of the nebular emission on our results separately in Appendix \ref{appendix:line_effect}.

We used the minimum $\chi^2$ method to obtain the best-fit model.
The $\chi^2$ was calculated as follows:
\begin{eqnarray}
  \chi^2 = \sum_{{\rm{filters\ }}i} \frac{(A\times f_{i,model} - f_{i,obs})^2}{\sigma_{i}^2},
  \label{eq:eq2}
\end{eqnarray}
where $A$, $f_{i,model}$, $f_{i,obs}$, and $\sigma_i$ are a normalization factor, flux of model spectrum, observed flux, and error of observed flux, respectively.
In order to take account of the variation of the IGM absorption among lines of sight \citep[e.g.,][]{Thomas_2017}, we added the half value of $f_{i,model}$ to the $\sigma_i$ in quadrature depending on the fraction of shorter wavelengths than $1216\rm{\AA}$ in the rest frame in each filter.

%%%%% chi^2
In order to remove the contamination in the $JHK_s$-band color-selected sample, we used the difference between the minimum $\chi^2$ values at the best-fit redshift and $z=2.39$ (hereafter we call $\Delta\chi^2_{z=2.39}$).
We defined the $\Delta\chi^2_{z=2.39}$ as follows:
\begin{eqnarray}
  \Delta\chi^2_{z=2.39} \equiv \chi^2_{z=2.39} - \chi^2_{min},
  \label{eq:eq3}
\end{eqnarray} 
where $\chi^2_{z=2.39}$ and $\chi^2_{min}$ are the minimum values of $\chi^2$ at $z=2.39$ and the best-fit redshift, respectively.
A higher value of $\Delta\chi^2_{z=2.39}$ means that the observed SEDs cannot be reproduced well by models at $z=2.39$ and there is high probability of other redshifts.
Therefore, we adopted the criterion of the $\Delta\chi^2_{z=2.39}$ to exclude contamination as follows:
\begin{eqnarray}
  \Delta\chi^2_{z=2.39} < 2.71.
  \label{eq:eq4}
\end{eqnarray}
This criterion means that 90\% confidence interval of the best-fit redshift includes $z=2.39$. 
The median full width of the 90\% confidence interval for the color-selected galaxies is $\Delta z\sim1.6$.
Finally, we selected 62 objects by using the $JHK_s$-color cuts and the $\Delta\chi^2_{z=2.39}$ criterion in the 53W002 field (hereafter $JHK_s$-selected galaxies).
In the next subsection, we demonstrate how the two-step selection works by using the multi-wavelength catalog from the COSMOS survey.

\subsection{Comparison Sample}\label{subsec:data_COSMOS}

In order to measure a number density excess of the $JHK_s$-selected galaxies in the 53W002 field and compare their statistical properties with field galaxies, we constructed a comparison sample in a general field from the COSMOS2015 catalog \citep{Laigle_2016} with the same selection method.
The COSMOS2015 catalog has multi-bands photometry from ultraviolet to far-infrared wavelength in a large area of about 2 deg$^2$, and also includes $B$, $V$, $i^{\prime}$, $J$, $H$, and $K_s$ bands, which are similar with those used in the 53W002 field.
We note that the $B$, $V$, and $i^{\prime}$-band data were taken by Suprime-Cam 
\citep{Taniguchi_2007}, while these $J$, $H$, and $K_s$-band data were taken by VIRCAM mounted on VISTA telescope \citep{McCracken_2012}.
Objects in this catalog were detected on the combined $zYJHK_s$-bands image.
The redshift is estimated by using SED fitting with about 30 bands photometric data, including the broadband and medium-band filters from near-UV to mid-infrared.
\citet{Laigle_2016} compared their estimated photometric redshifts with spectroscopic redshifts and calculated the dispersion of $(z_{photo}-z_{spec})/(1+z_{spec})$ (hereafter $\sigma_z$) and the fraction of catastrophic failure, namely, $|z_{photo}-z_{spec}|$/$(1+z_{spec})>0.15$ (hereafter $\eta$).
They reported that galaxies of zCOSMOS faint sample at $z=1.5$--$2.5$ show $(\sigma_z,\eta)=(0.0032,0.08)$, while those of 80 galaxies from MOSDEF survey \citep{Kriek_2015} are $(0.042, 0.10)$.
Furthermore, several quiescent galaxies at $z\sim1.2$--$2.5$ show low $\sigma$ of 0.017--0.069 with no catastrophic failure (see \citealp{Laigle_2016} for more details).

In this study, we used the $B$, $V$, $i^{\prime}$, $J$, $H$, and $K_s$-band fluxes and their errors measured with a $3^{\prime\prime}$ diameter aperture.
We selected objects with {\tt{FLAG\_COS=1}} and {\tt{FLAG\_DEEP=1}} to use objects in both the COSMOS field and Ultradeep stripes of the UltraVISTA survey.
It should be noted that objects in ultra-deep stripe 4, the most western one, were excluded because the area overlapped with the COSMOS field is too small to sample a random field of view of about $10^{\prime}\times10^{\prime}$ (see Section \ref{sec:result_numden}).
Since the different depths of the data could affect the sample selection, we performed Monte-Carlo simulations where random shifts were added to the $J$, $H$, and $K_s$-band fluxes of the COSMOS2015 catalog to match the flux errors with those of the 53W002 data as follows:
\begin{eqnarray}
  F = F_{0} + R_{gauss}\times\sqrt{\sigma_{\rm{53W002}}^2-\sigma_{\rm{COSMOS}}^2},
  \label{eq:eq5}
\end{eqnarray}
where $F$, $F_0$, $R_{gauss}$, $\sigma_{\rm{53W002}}$, and $\sigma_{\rm{COSMOS}}$ are a simulated flux, original flux in the COSMOS2015 catalog, random numbers that follow a normal distribution with a mean of zero and a standard deviation of one, the median background error measured in $1\farcs6$ aperture in the 53W002 field, and that in $3\farcs0$ aperture in the COSMOS field, respectively.
We note that the procedure mentioned above was not applied for the optical data, because the optical data in the 53W002 field have the same depth or slightly deeper than that in the COSMOS field.
After convolving the NIR fluxes, we applied the total $K_s$-band magnitude cut of $K_s<23.3$, which is the same limiting magnitude as in the 53W002 field. 
We ran SExtractor on the $K_s$-band image of the UltraVISTA Dr2 and used the Kron magnitudes from SExtractor as the total magnitudes for the objects in the COSMOS field.
Then we corrected the total $K_s$-band magnitude for the different depths between the COSMOS and 53W002 data by adding the same random error as the $3^{\prime\prime}$ aperture photometry in the $K_s$ band (i.e., the second term of the equation (\ref{eq:eq5})).

We then selected galaxies at $z\sim2.39$ with the same selection method as in the 53W002 field (i.e., the $JHK_s$-color and $\Delta\chi^2_{z=2.39}$ selections). 
1182 objects were selected by the $JHK_s$-color cuts. 
We note that objects with $H>23.8$, which is the brightest $3\sigma$ limiting magnitude of the $H$ band in the 53W002 field, were removed to match the selection criterion to that for the 53W002 sample. 
After that, we applied the $\Delta\chi^2_{z=2.39}$ criterion of the equation (\ref{eq:eq4}) and selected 896 galaxies.

\begin{figure}[ht!]
  \includegraphics[width=8.5cm]{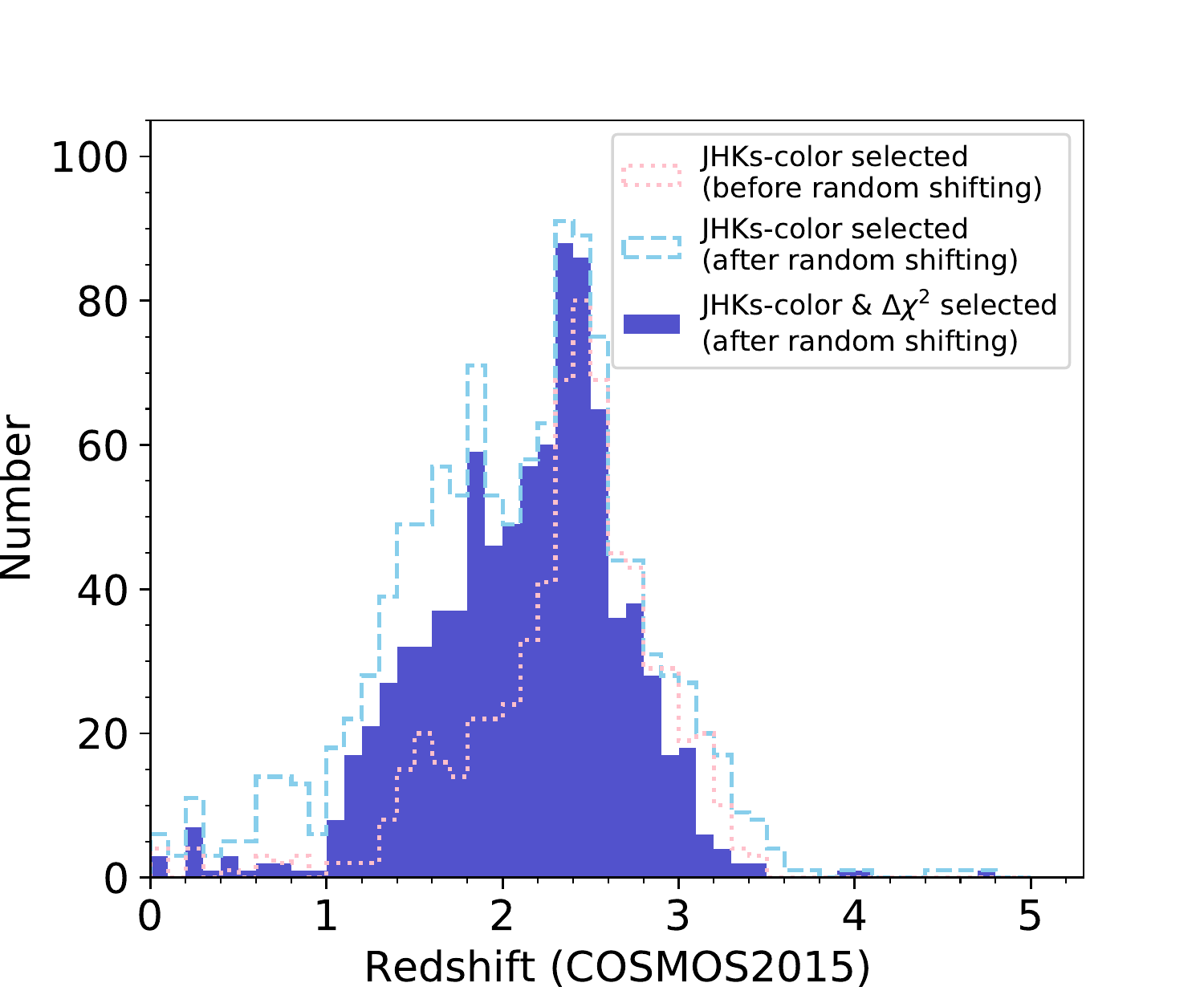}
  \caption{Redshift distribution of the selected objects in the COSMOS field.
  The photometric redshifts from the COSMOS2015 catalog \citep{Laigle_2016} are used.
  The blue solid histogram shows the distribution for the $JHK_s$-selected galaxies in the COSMOS field. 
  The red dotted and blue dashed lines represent objects selected by the $JHK_s$-bands color criteria before and after the addition of the random error to the NIR fluxes (see text), respectively.}
  \label{fig:fig3}
\end{figure}
Since the photometric redshifts from the COSMOS2015 catalog, which are estimated with the 30-bands photometries, have relatively high accuracy as mentioned above, we can use it to check the redshift distribution of the selected objects and the effects of the difference in the depth of the NIR data and the additional $\Delta\chi^2_{z=2.39}$ criterion.
Figure \ref{fig:fig3} shows the distribution of the photometric redshifts from the COSMOS2015 catalog for the $JHK_s$-selected objects in the COSMOS field (solid histogram).
For comparison, we also show the same distribution for those objects selected only by the $JHK_s$-color cuts (without the $\Delta\chi^2_{z=2.39}$) before and after the addition of the random error to the NIR fluxes (red dotted and blue dashed lines in the figure).
Note that the addition of the random error affects the $JHK_s$-color cuts and the calculation of $\Delta\chi^2_{z=2.39}$ from the $BVi^{\prime}JHK_s$-bands photometries, but the photometric redshifts in the figure are the values in the COSMOS2015 catalog in all the cases (i.e., not affected by the additional noise).
While the shallower depth of the NIR data (the additional random noise in the fluxes) leads to a significant increase of the contamination from foreground in the $JHK_s$-bands color selection, the addition of the $\Delta\chi^2_{z=2.39}$ criterion mitigates the contamination without losing many objects around $z\sim2.4$.
The resulting redshift distribution shows a peak around $z\sim2.4$, while 38\% of the $JHK_s$-selected objects have $z_{photo}<2$ in the COSMOS2015 catalog.

\subsection{Detection Completeness}\label{subsec:completeness}
While we took the difference in the photometric errors between the 53W002 and COSMOS fields into account in the construction of the comparison sample, the detection completeness in the $K_s$ band can be different between the two fields.
Therefore, we carried out simulations to investigate the difference in the detection completeness.
We here briefly describe the simulations and their results and refer the reader to Appendix \ref{appendix:comple} for more details.

In the simulations, we aimed to measure what fraction of $JHK_s$-selected galaxies from the comparison sample can be detected on the $K_s$-band image of the 53W002 field as a function of $K_s$-band magnitude.
We did this by adding artificial objects with similar surface brightness as the $JHK_s$-selected galaxies detected in the COSMOS field to the $K_s$-band data of the 53W002 field, and performing the source detection on the $K_s$-band image to check whether the added objects are detected or not.
Since the PSF sizes of the $K_s$-band data are different between the 53W002 and COSMOS fields, we at first estimated intrinsic sizes of the $JHK_s$-selected galaxies in the COSMOS field and used them to make the artificial objects in the main simulation.
In order to estimate the intrinsic sizes, we measured apparent half-light radii of the $JHK_s$-selected galaxies on the $K_s$-band image of the UltraVISTA DR2 with SExtractor and converted them to the intrinsic sizes by using the relation between the apparent and intrinsic sizes for these objects derived from the simulations adding artificial objects to the same $K_s$-band data with {\tt{IRAF/MKOBJECTS}} task.

\begin{figure}[ht!]
  \centering
  \includegraphics[width=8cm]{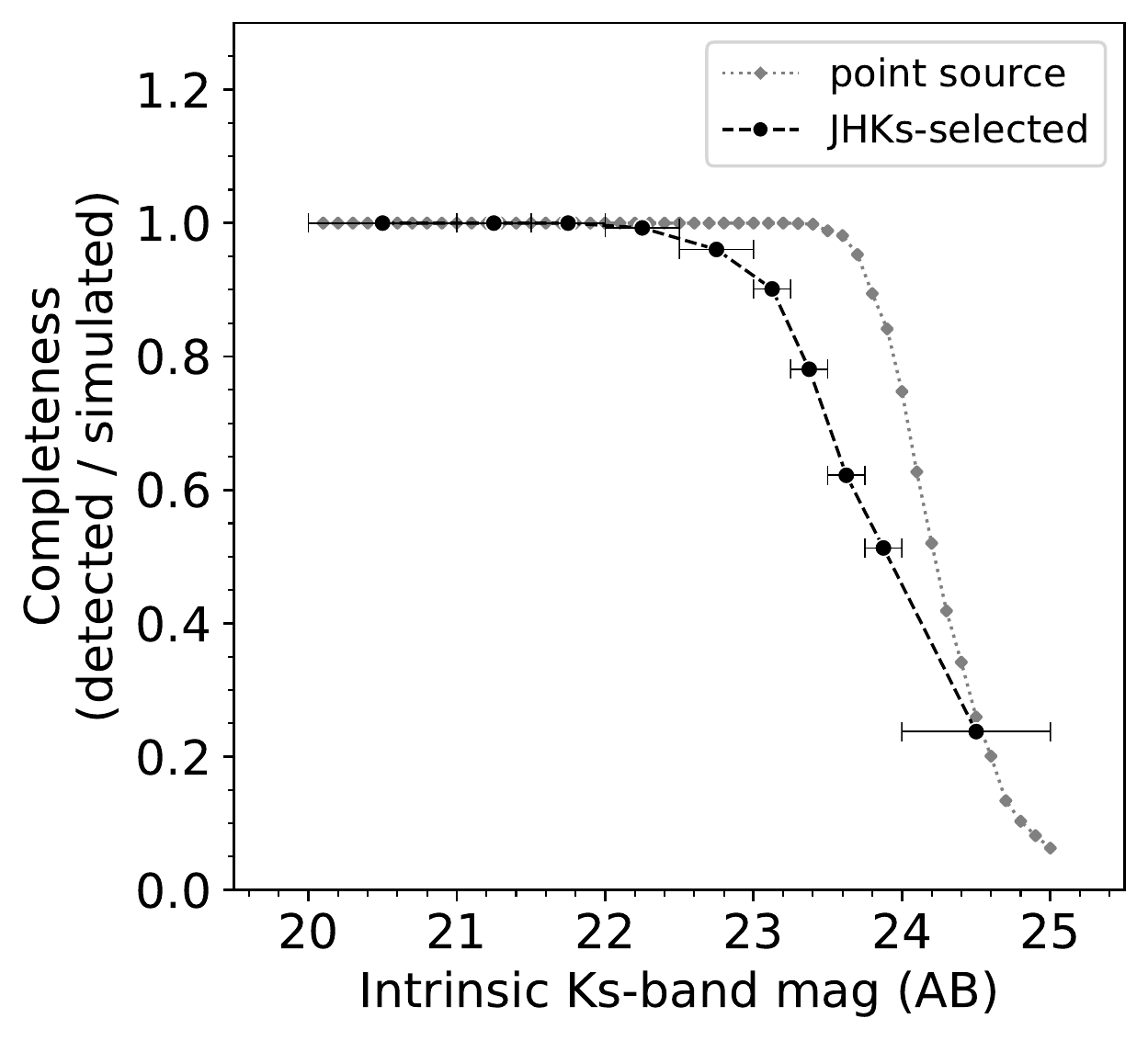}
  \caption{The detection completeness in the 53W002 field. The black circles mean the detection completeness with the same size distribution as those detected in the COSMOS field. The grey diamonds are the same as the black circles, but for the point source.}
  \label{fig:fig4}
\end{figure}

We then convolved artificial objects that have the estimated intrinsic sizes of the $JHK_s$-selected galaxies in the COSMOS field with the PSF of the $K_s$-band data of the 53W002 field and added them to random positions in the $K_s$-band sky image described in Section \ref{subsec:detection} with {\tt{IRAF/MKOBJECTS}}.
The same source detection as in Section \ref{subsec:detection} was performed on the $K_s$-band images.
We repeatedly performed the simulations and measured the fraction of detected artificial objects as a function of $K_s$-band magnitude.
Figure \ref{fig:fig4} shows the estimated detection completeness for the $JHK_s$-selected galaxies in the 53W002 field.
We also show that for point sources for comparison in the figure.
The completeness for the $JHK_s$-selected galaxies begins to decrease at $K_s\sim22.5$ and become $\sim90\%$ at $K_s\sim23.0$ and $\sim80\%$ at $K_s\sim23.3$.
On the other hand, that for point sources is $\sim100\%$ at $K_s<23.3$, which indicates that the effect of the sizes of these galaxies is non-negligible.
In Section \ref{sec:results}, we corrected number density of the comparison sample by multiplying it with the detection completeness for the $JHK_s$-selected galaxies for fair comparison.

\subsection{Physical Properties}\label{subsec:phys_prop}
Assuming that the selected member candidates are at $z=2.39$, we estimated their physical properties such as stellar mass ($M_s$), star formation rate (SFR), dust extinction, and so on, from the results of the SED fitting described in Section \ref{subsec:sample}.
We corrected $M_s$ and SFR for the offset between the total and aperture magnitudes in the $K_s$ band.
We calculated SFR averaged over the past 100 Myr from the best-fit star formation history (SFH) and use it in this study rather than the instantaneous SFR, which tends to have large uncertainty in our SED fitting setup (assumed parametric forms of SFH).
We estimated the specific SFR (sSFR) by dividing the obtained star formation rate (averaged over the past 100 Myr) by the stellar mass.
As an indicator of the stellar age, we estimated the mass-weighted age as follows:
\begin{eqnarray}
  {\rm{mass\mathchar`-weighted\ age}} = \frac{\int^{age}_{0} (age-t)\times {\rm{SFR}}(t) dt}{\int^{age}_{0} {\rm{SFR}}(t) dt},
  \label{eq:eq6}
\end{eqnarray}
where $age$ and SFR$(t)$ are the age of the best-fit model spectrum and SFR as a function of time elapsed from the formation epoch (the beginning of star formation), respectively.
We note that all stars formed since the formation epoch are included in the calculation.

%%%%%%%%%%%%%%%%%%%%%%%%%%%%%%%%%%%%%%%%%%%%%%%%%%%%%%%%%%%%%%%%%%
%%%%%%%                                                    %%%%%%%
%%%%%%%                                                    %%%%%%%
%%%%%%%                       Result                       %%%%%%%
%%%%%%%                                                    %%%%%%%
%%%%%%%                                                    %%%%%%%
%%%%%%%%%%%%%%%%%%%%%%%%%%%%%%%%%%%%%%%%%%%%%%%%%%%%%%%%%%%%%%%%%%
\section{Results} \label{sec:results}
\subsection{Number Density Excess in the 53W002 Field}\label{sec:result_numden}
\begin{figure*}[ht!]
  \includegraphics[width=17cm]{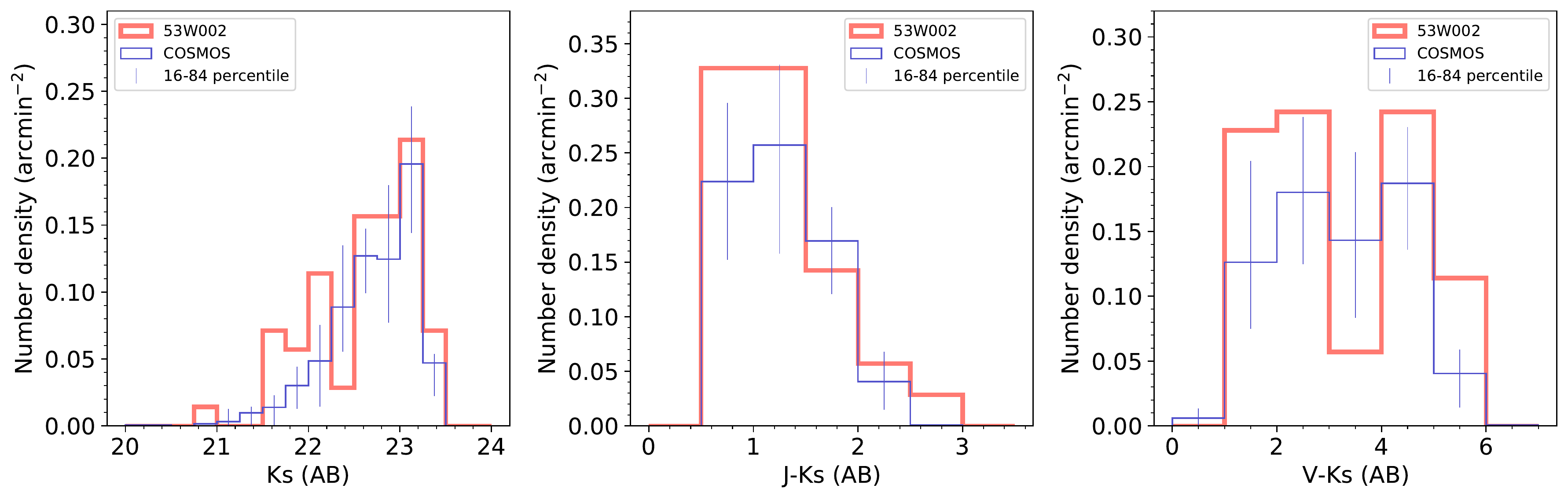}
  \caption{Left: $K_s$-band number counts of the $JHK_s$-selected galaxies in the 53W002 (red) and COSMOS (blue) fields. Error bars represent the 16 and 84 percentiles of the result in 200 random fields of view of $10^\prime\times10^\prime$ in the COSMOS field (see text). 
  The number density in the COSMOS field and its error are corrected for the difference in the detection completeness between the two fields, and each object is multiplied by the completeness for $JHK_s$-selected galaxies in the 53W002 field depending on its $K_s$-band magnitude in the calculation (see text).
  Middle: $J-K_s$ color distribution of the selected galaxies in the 53W002 and COSMOS fields. The error bars are the same as in the left panel. Right: The same as the middle panel but for $V-K_s$ color.}
  \label{fig:fig5}
\end{figure*}
\begin{figure*}[ht!]
  \includegraphics[width=17cm]{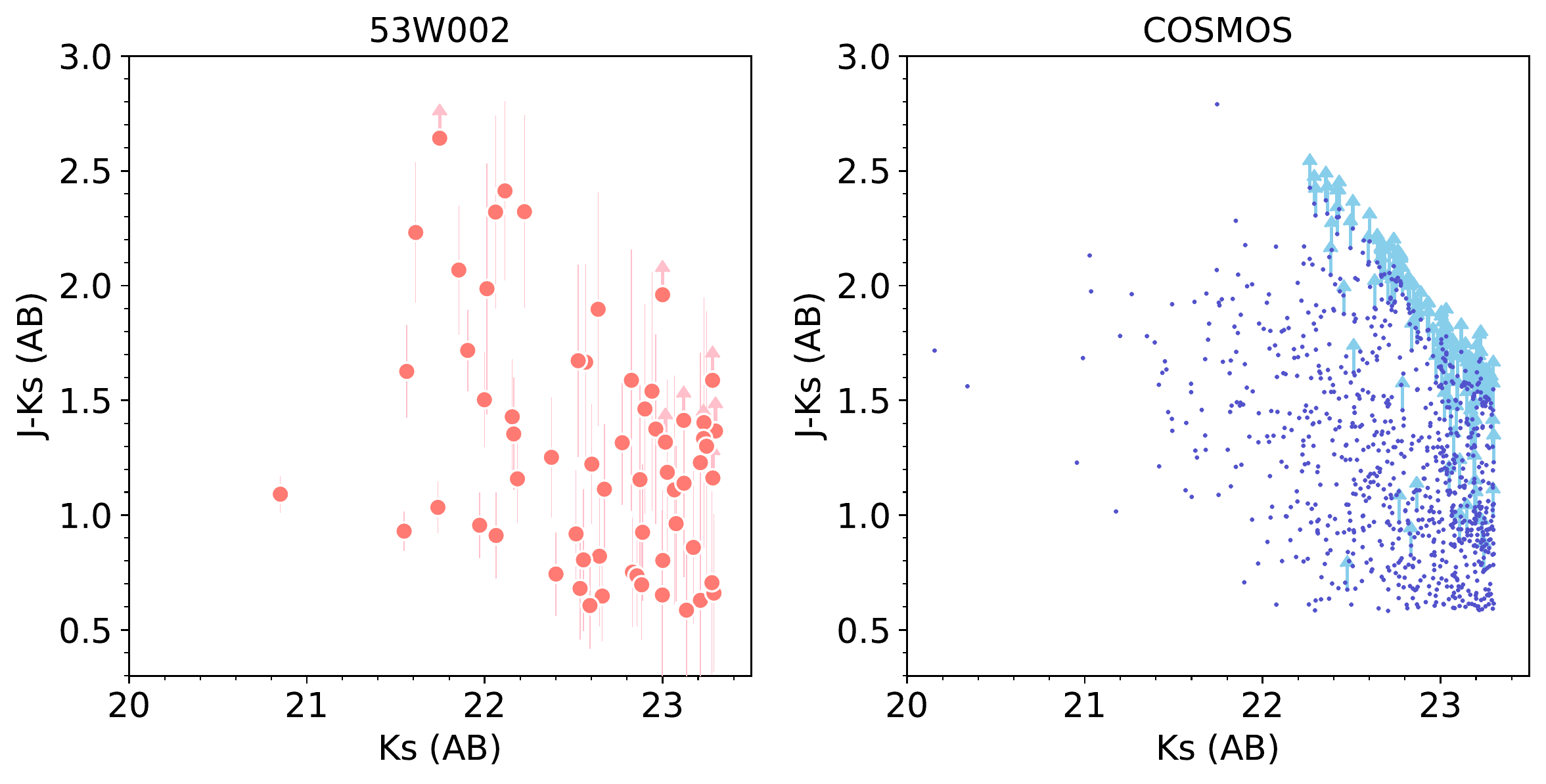}
  \caption{$J-K_s$ vs.~$K_s$ color-magnitude diagram for the $JHK_s$-selected galaxies in the 53W002 (left panel) and COSMOS (right) fields. The symbols with allows represent the lower limit of $J-K_s$ color. In the right panel, the error bars are omitted for clarity.}
  \label{fig:fig6}
\end{figure*}
\begin{deluxetable*}{c c ccc cc}
  \tablecaption{Number densities for the $JHK_s$-selected galaxies with the different magnitude and color cuts.\label{tab:table3}}
  \tablewidth{0pt}
  \tablehead{
    \colhead{} &
    \multicolumn{6}{c}{Number density (arcmin$^{-2}$)}
    \\ \cline{2-7}
    \colhead{Condition} &
    \colhead{$K_s<23.3$} &
    \colhead{$K_s<22.25$} & \colhead{$J-K_s>2$}& \colhead{$V-K_s>4$} & 
    \colhead{\begin{tabular}{c}$J-K_s>2$\\\&\\$K_s<22.25$\end{tabular}}  & 
    \colhead{\begin{tabular}{c}$V-K_s>4$\\\&\\$K_s<22.25$\end{tabular}}
  }
  %\decimalcolnumbers
  \startdata
  53W002 field & 0.88 & 0.27 & 0.085 & 0.36 & 0.085 & 0.19
  \smallskip
  \\
  COSMOS field\tablenotemark{a} & 0.69 & 0.11 & 0.041 & 0.23 & 0.009 & 0.066
  \smallskip
  \\\hline
  Reproduced rate\tablenotemark{b} & 12.5\% & 0.0\% & 6.0\% & 0.0\% & 0.0\% & 0.0\%
  \enddata
  \tablenotetext{a}{The number densities of those galaxies were corrected for the difference in the $K_s$-band completeness between the 53W002 and COSMOS fields.}
  \tablenotetext{b}{The fraction of random fields of view of $10^\prime\times10^\prime$ in the COSMOS field with a higher number density than that in the 53W002 field.}
  %\tablecomments{}
\end{deluxetable*}
\begin{figure*}[ht!]
  \includegraphics[width=17cm]{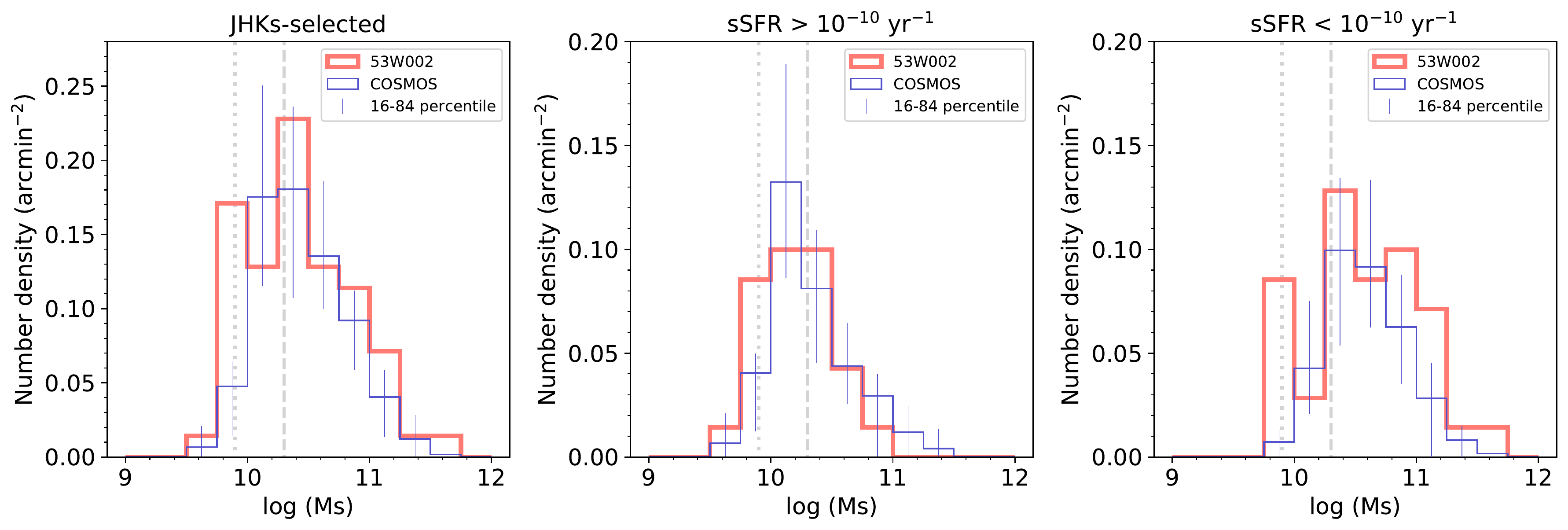}
  \caption{Left: Stellar mass distribution of the $JHK_s$-selected galaxies in the 53W002 (red) and COSMOS (blue) fields. The number density in the COSMOS field was corrected for the difference in the detection completeness between the two fields as in Figure \ref{fig:fig5}. The error bars are the same as in Figure \ref{fig:fig5}. The dotted and dashed lines indicate stellar mass completeness limits (see Appendix \ref{appendix:Mslim}) at mass-weighted age = $10^{8}$ and $10^9$ yr, respectively. Middle: The same as the left panel but for those with sSFR $>10^{-10}\ \rm{yr^{-1}}$. Right: The same as the left panel but for those with sSFR $<10^{-10}\ \rm{yr^{-1}}$.}
  \label{fig:fig7}
\end{figure*}

\vspace{-8mm}
In this subsection, we compare the number density of the $JHK_s$-selected galaxies in the 53W002 field with that in the COSMOS field.
By the $JHK_s$-color cuts and $\Delta\chi^2_{z=2.39}$ selection, we selected 62 objects with $K_s<23.3$ in the 53W002 field over 70.2 arcmin$^{2}$.
On the other hand, 896 objects were selected in the COSMOS field over 1230 arcmin$^{2}$.
In order to consider the difference in the detection completeness for the $JHK_s$-selected galaxies between the 53W002 and COSMOS data described in Section \ref{subsec:completeness}, we corrected the number of objects in the COSMOS field by multiplying each object by the detection completeness in the 53W002 field depending on its $K_s$-band magnitude.
The corrected number of objects in the COSMOS field became 850.7.
These results correspond to the number densities of 0.88 and 0.69 arcmin$^{-2}$ in the 53W002 and COSMOS fields, respectively.
In order to estimate the field-to-field variation for the number density of the $JHK_s$-selected galaxies in an area of the 53W002 field, we performed a Monte-Carlo simulation with the comparison sample.
We set $10^\prime\times10^\prime$ fields of view at random positions in the COSMOS field, and calculated the number density in each field, which has a similar effective area with the 53W002 field after excluding masked regions.
We repeated 200 such procedures.
As a result, the median value and standard deviation of the number density in the random fields are 0.65 and 0.19 arcmin$^{-2}$, and 12.5\% of the Monte-Carlo realizations in the COSMOS field show a higher number density than that in the 53W002 field.

The left panel of Figure \ref{fig:fig5} shows $K_s$-band number counts of the $JHK_s$-selected galaxies in the 53W002 and COSMOS fields.
The error bars show the 16 and 84 percentiles of the number density estimated from the same Monte-Carlo simulation in the COSMOS field.
The number density of the $JHK_s$-selected galaxies with $K_s < 22.25$ in the 53W002 field is about 2.5 times higher than that in the COSMOS field, and no random field of view in the simulation shows number density higher than the 53W002 field.
While there is a possible deficit in the $K_s=22.25$--$22.5$ bin, a marginal excess can be seen at $K_s>22.5$, where the number density in each bin is consistent with the COSMOS field within the uncertainty except for the faintest bin.

The middle and right panels of Figure \ref{fig:fig5} show the color distributions of the $J-K_s$ and $V-K_s$, respectively.
There is a density excess of red objects with $J-K_s>2$ in the 53W002 field.
The fraction of random fields of view with a higher density of such red objects than that in the 53W002 field is 6.0\%.
In the right panel, red objects with $V-K_s>4$ also show a density excess, and none of the random fields of view in the COSMOS field reproduces the number density in the 53W002 field.
We also focused on such red objects with $K_s<22.25$, where the number density excess is seen in the number count.
These bright and red objects with $K_s<22.25$ and $J-K_s>2$ ($V-K_s>4$) in the 53W002 field show a number density about nine (three) times higher than those in the COSMOS field.
None of random fields of view in the COSMOS field shows a higher density of these bright and red objects than the 53W002 field.
The number densities for these samples are summarized in Table \ref{tab:table3}.

We note that the $JHK_s$-selected galaxies with blue colors of $J-K_s<1.5$ or $V-K_s<3$ also show a number density excess in the 53W002 field.
15.0 and 11.0\% of the Monte-Carlo realizations in the COSMOS field show a higher number density than the 53W002 field for those with $J-K_s<1.5$ and $V-K_s<3$, respectively.

Figure \ref{fig:fig6} shows the color-magnitude diagrams in the 53W002 and COSMOS fields.
The $J-K_s$ colors of these bright and red objects in the 53W002 field show a relatively large scatter rather than a tight ``red sequence'' found in lower-redshift clusters \citep{Kodama_2007}.

\subsection{Physical Properties}\label{subsec:result_prop}
\begin{figure*}[p]
  \centering
  \includegraphics[width=18.5cm]{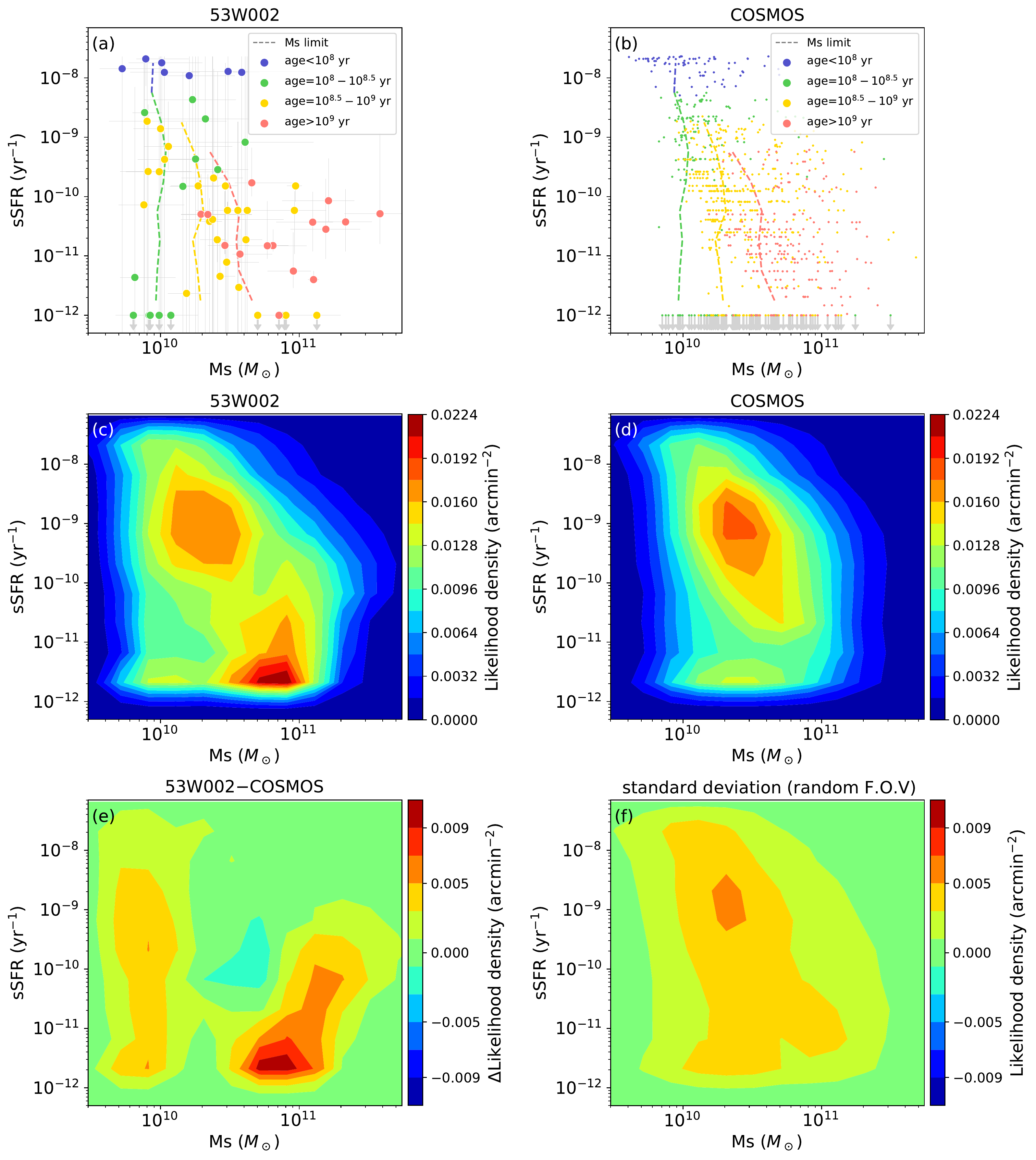}
  \caption{(a): sSFR vs.~$M_s$ for the $JHK_s$-selected galaxies in the 53W002 field. Symbols with arrows mean sSFR $<10^{-12}\ \rm{yr^{-1}}$. 
  The different colors show different mass-weighted ages of the selected objects. 
  The dashed lines indicate the stellar mass completeness limits, above which all galaxies are brighter than $K_s=23.3$, as a function of sSFR for the different mass-weighted ages (Appendix \ref{appendix:Mslim}). 
  (b): The same as the panel (a) but in the COSMOS field. Error bars are omitted for clarity. 
  (c): Probability density distribution in sSFR vs.~$M_s$ for the selected objects in the 53W002 field. 
  (d): The same as the panel (c) but in the COSMOS field. The probability density was corrected for the difference in the detection completeness between the 53W002 and COSMOS fields. 
  (e): Excess map of the probability density distribution in the 53W002 field, which represents the differences in the probability density between the 53W002 and COSMOS fields. 
  (f): Field variance of the probability density distribution for random fields of view of $10^\prime\times10^\prime$ in the COSMOS field.}
  \label{fig:fig8}
\end{figure*}
\begin{figure*}[h]
  \centering
  \includegraphics[width=18cm]{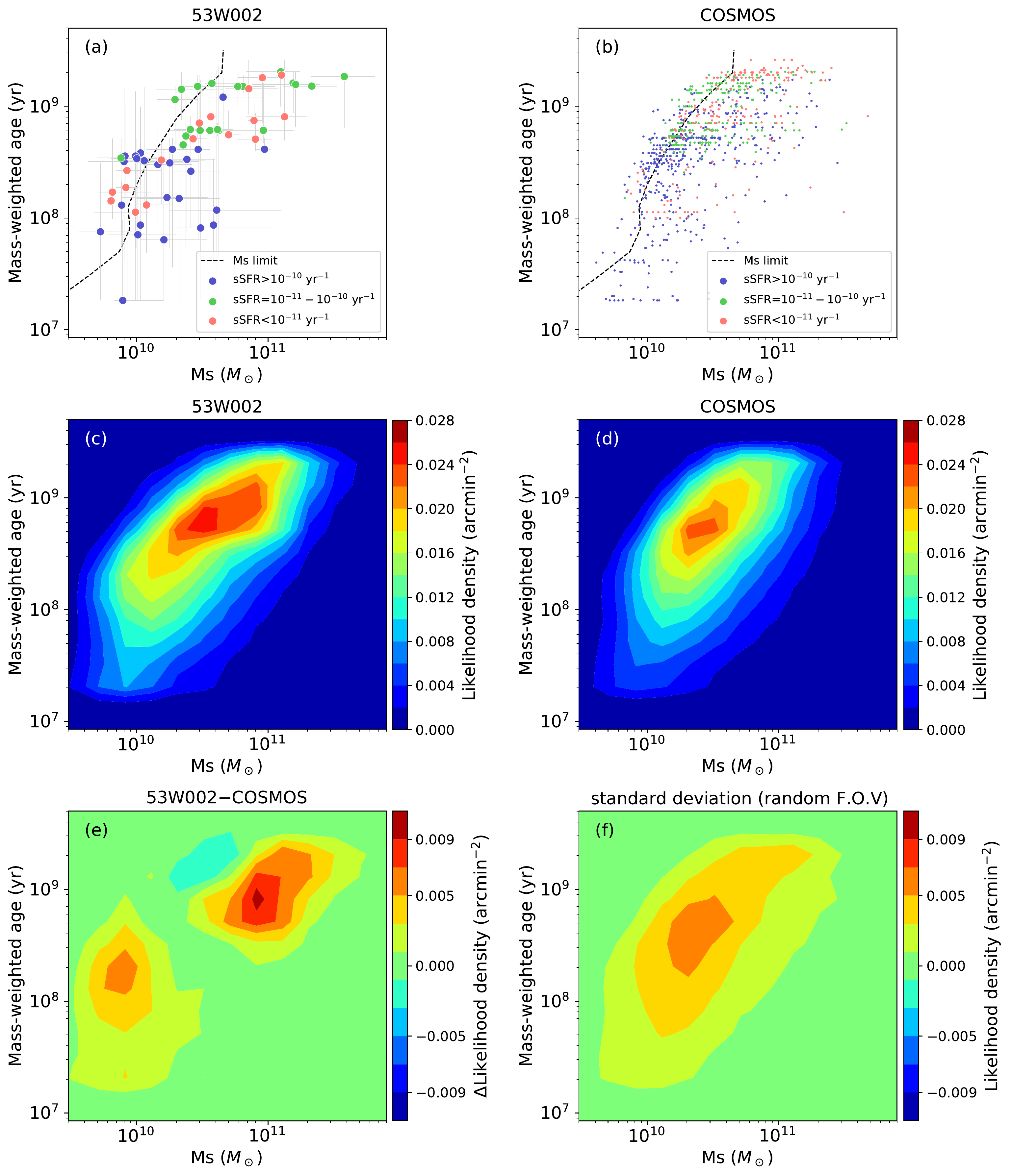}
  \caption{
    The same as Figure \ref{fig:fig8}, but for mass-weighted age vs.~$M_s$. 
    In the panel (a) and (b), different symbol colors indicate the different sSFRs of the $JHK_s$-selected galaxies.
    The dashed line shows the stellar mass completeness limit as a function of mass-weighted age (Appendix \ref{appendix:Mslim}).
  }
  \label{fig:fig9}
\end{figure*}
\begin{figure*}[ht!]
  \centering
  \includegraphics[width=15cm]{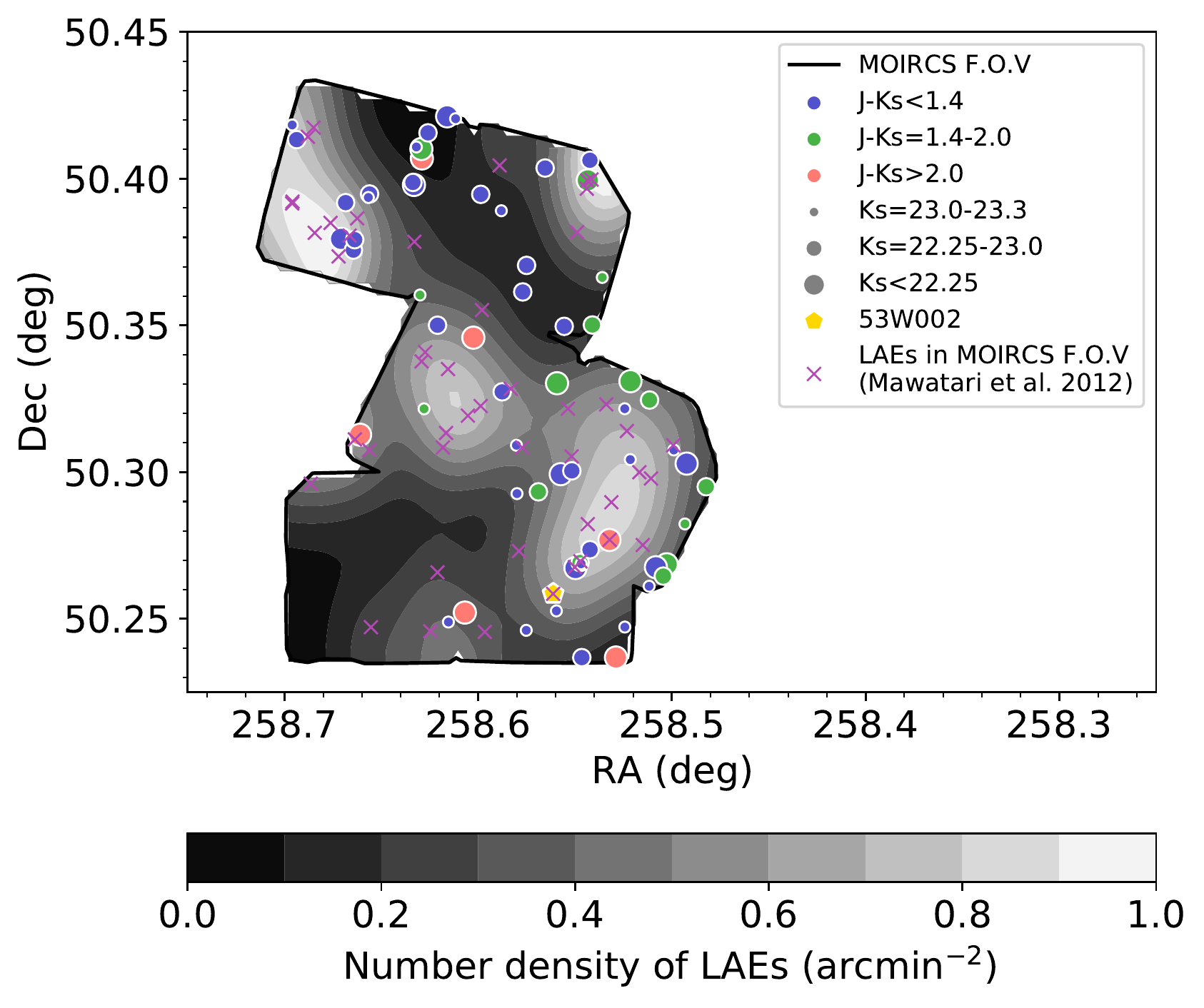}
  \caption{Spatial distribution of the $JHK_s$-selected galaxies (circles) and LAEs (crosses) in our survey field.
  The different colors of the circles represent objects with different $J-K_s$ colors, while the different sizes of the symbols show different $K_s$-band magnitudes.
  The yellow pentagon shows the radio galaxy 53W002.
  The grey-scale contour represents the number density map of LAEs in the field smoothed by a Gaussian kernel with $\sigma=1\farcm0$, which corresponds to $\sim1.7$ cMpc at $z=2.39$.
  }
  \label{fig:fig10}
\end{figure*}

The left panel of Figure \ref{fig:fig7} shows the stellar mass distribution of the $JHK_s$-selected galaxies. 
The red and blue lines indicate results in the 53W002 and COSMOS fields, respectively.
We also show stellar mass completeness limit, above which all galaxies are brighter than the magnitude limit of $K_s=23.3$ for a given mass-weighted stellar age (see Appendix \ref{appendix:Mslim} for detail).
The number density in the 53W002 field is higher than that in the COSMOS field at $M_s\gtrsim10^{11}\ M_\odot$ and $M_s<10^{10}\ M_\odot$.
The middle and right panels of Figure \ref{fig:fig7} show the stellar mass distributions for the $JHK_s$-selected galaxies with sSFR $>10^{-10}\ \rm{yr^{-1}}$ and sSFR $<10^{-10}\ \rm{yr^{-1}}$, respectively.
From these panels, one can see that those with sSFR $<10^{-10}\ \rm{yr^{-1}}$ mainly contribute to the number density excess in the 53W002 field.
The number density of the $JHK_s$-selected galaxies with sSFR $<10^{-10}\ \rm{yr^{-1}}$ is about 1.5 times higher than those in the COSMOS field, while that of the galaxies with sSFR $>10^{-10}\ \rm{yr^{-1}}$ is comparable to the comparison sample.
For the $JHK_s$-selected galaxies with sSFR $<10^{-10}\ \rm{yr^{-1}}$ (sSFR $>10^{-10}\ \rm{yr^{-1}}$), 5.0\% (34.0\%) of the Monte-Carlo realizations in the COSMOS field show a higher number density than that in the 53W002 field.
In particular, the number density excess of those with sSFR $<10^{-10}\ \rm{yr^{-1}}$ is significant at $M_s>10^{10.75}\ M_\odot$ and $M_s=10^{9.75}$--$10^{10}\ M_\odot$, where no Monte-Carlo realization shows a higher number density than that in the 53W002 field.
On the other hand, galaxies with higher sSFRs show a density excess only at $M_s<10^{10}\ M_\odot$, and a marginal deficit at $M_s>10^{10}\ M_\odot$.

We also investigated the relation between sSFR and $M_s$.
Panels (a) and (b) of Figure \ref{fig:fig8} show sSFR vs.~$M_s$ in the 53W002 and COSMOS fields.
The different colors indicate the $JHK_s$-selected galaxies with different mass-weighted ages.
We also show the stellar mass completeness limit as a function of sSFR for each mass-weighted age (see Appendix \ref{appendix:Mslim}).
In the 53W002 field, 83\% of those with $M_s>10^{10.5}\ M_\odot$ show sSFR $<10^{-10}\ \rm{yr^{-1}}$, while 84\% of those with sSFR $>10^{-10}\ \rm{yr^{-1}}$ have $M_s<10^{10.5}\ M_\odot$.
Low-mass galaxies with $M_s\sim10^{10}\ M_\odot$ show a variety of sSFR from $\lesssim10^{-11}\ \rm{yr^{-1}}$ to $\gtrsim10^{-8}\ \rm{yr^{-1}}$, although the uncertainty in their sSFRs is relatively large.
One can see a group of massive galaxies with $M_s\gtrsim10^{11}\ M_\odot$ and sSFR $\sim10^{-10.5}-10^{-10}\ \rm{yr^{-1}}$ in the 53W002 field.
Their sSFRs suggest that their star formation has not yet stopped completely.
Most of them have older mass-weighted ages of $>10^{9}$ yr.

In order to directly compare the distribution in the sSFR vs.~$M_s$ plane and subtract the contribution of field galaxies, we made a probability density distribution in this plane.
First, we divided this plane into rectangular grids with an equal spacing in $\log M_s$ and $\log$ sSFR, and searched a minimum $\chi^2$ value in each grid in the SED fitting for each object after replacing data with sSFR $<10^{-12}\ \rm{yr^{-1}}$ to sSFR $=10^{-12}\ \rm{yr^{-1}}$.
Next, we converted $\chi^2_{min}$ for each grid into a probability of $P\propto e^{-\chi^2_{min}/2}$, which was normalized so that the summation over all grids equals to 1.
Finally, we stacked the probability of all sample galaxies in each grid and then divided it by the survey area to obtain the probability density distribution.
Panels (c) and (d) of Figure \ref{fig:fig8} show the probability density distributions in the 53W002 and COSMOS fields, respectively.
We subtracted the probability density distribution in the COSMOS field from that in the 53W002 field to derive probability density excess in the 53W002 field (panel (e) of Figure \ref{fig:fig8}).
There is a clear number density excess at $M_s\sim5\times10^{10}$--$10^{11}\ M_\odot$ and sSFR $<10^{-11}\ \rm{yr^{-1}}$ in the 53W002 field.
One can see also excesses at $M_s>10^{11}\ M_\odot$ and $M_s<10^{10}\ M_\odot$ in the probability density excess map.
The panel (f) of Figure \ref{fig:fig8} shows the variation of the probability density distribution in random fields of view in the COSMOS field.
The dispersion among random fields is about 0.003--0.007 arcmin$^{-2}$, and the excesses mentioned above are larger than the field-to-field variation, especially for quiescent galaxies with $M_s=5\times10^{10}$--$10^{11}\ M_\odot$ and sSFR $<10^{-11}\ \rm{yr^{-1}}$.
We note that the observed SEDs of the low-mass galaxies can be affected by strong emission lines such as [OIII]$\lambda\lambda4959,5007$ and H$\alpha$, which enter into $H$ and $K_s$ bands at $z\sim2.39$.
In fact, some of these galaxies can also be fitted well with models with high sSFR of $\gtrsim10^{-8}\ \rm{yr^{-1}}$ and strong emission lines (Appendix \ref{appendix:line_effect}). 
However, the trend of the excess of low-mass galaxies with $M_s<10^{10}\ M_\odot$ would not change significantly even if we include the nebular emission in the SED fitting as shown in the appendix.
On the other hand, galaxies with $M_s=1$--$5\times10^{10}\ M_\odot$ and sSFR $=10^{-10.5}$--$10^{-9}$ yr$^{-1}$ in the 53W002 field have a smaller number density than in the COSMOS field.
Although the deficit may be within the field-to-field variation, it is also possible that there are few such intermediate-mass star-forming galaxies in this protocluster.
\begin{figure*}[ht!]
  \centering
  \includegraphics[width=18cm]{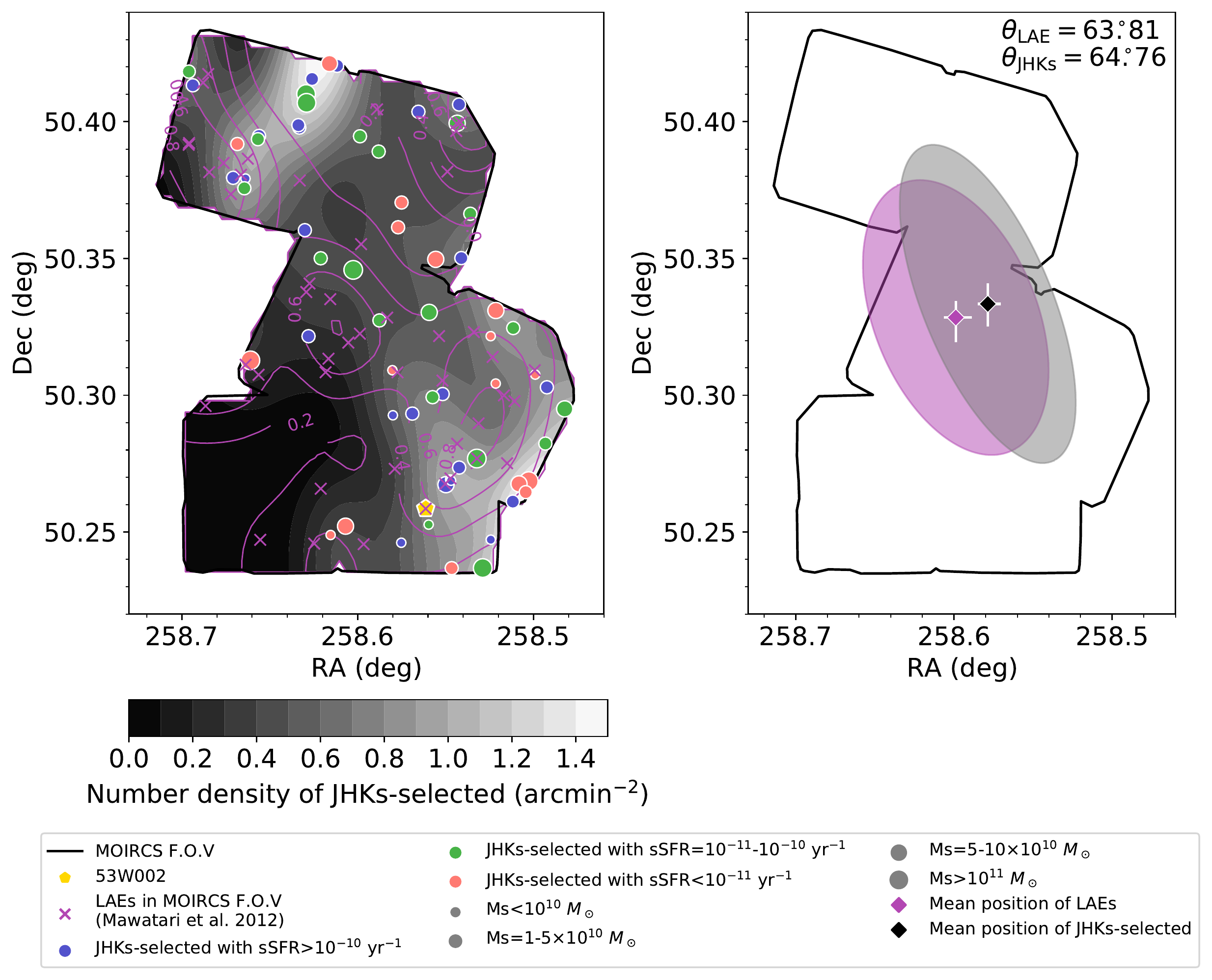}
  \caption{Left: The same as Figure \ref{fig:fig10}, but different symbol colors and sizes represent the different sSFRs and stellar masses of the $JHK_s$-selected galaxies.
  The grey scale contour shows the number density of the $JHK_s$-selected galaxies smoothed by a Gaussian kernel with $\sigma=1\farcm0$.
  The purple contour shows the number density of LAEs in the field.
  Right: Ellipses with the mean positions and semi-major and minor axes of the second order moments for the $JHK_s$-selected galaxies and LAEs (see text). The black and purple diamonds indicate the mean positions of the $JHK_s$-selected galaxies and LAEs, respectively. The white error bars represent 68\% confidence range estimated with the bootstrap method. The position angles for the samples are also shown in the panel.
  }
  \label{fig:fig11}
\end{figure*}
\begin{figure*}[ht!]
  \centering
  \includegraphics[width=18cm]{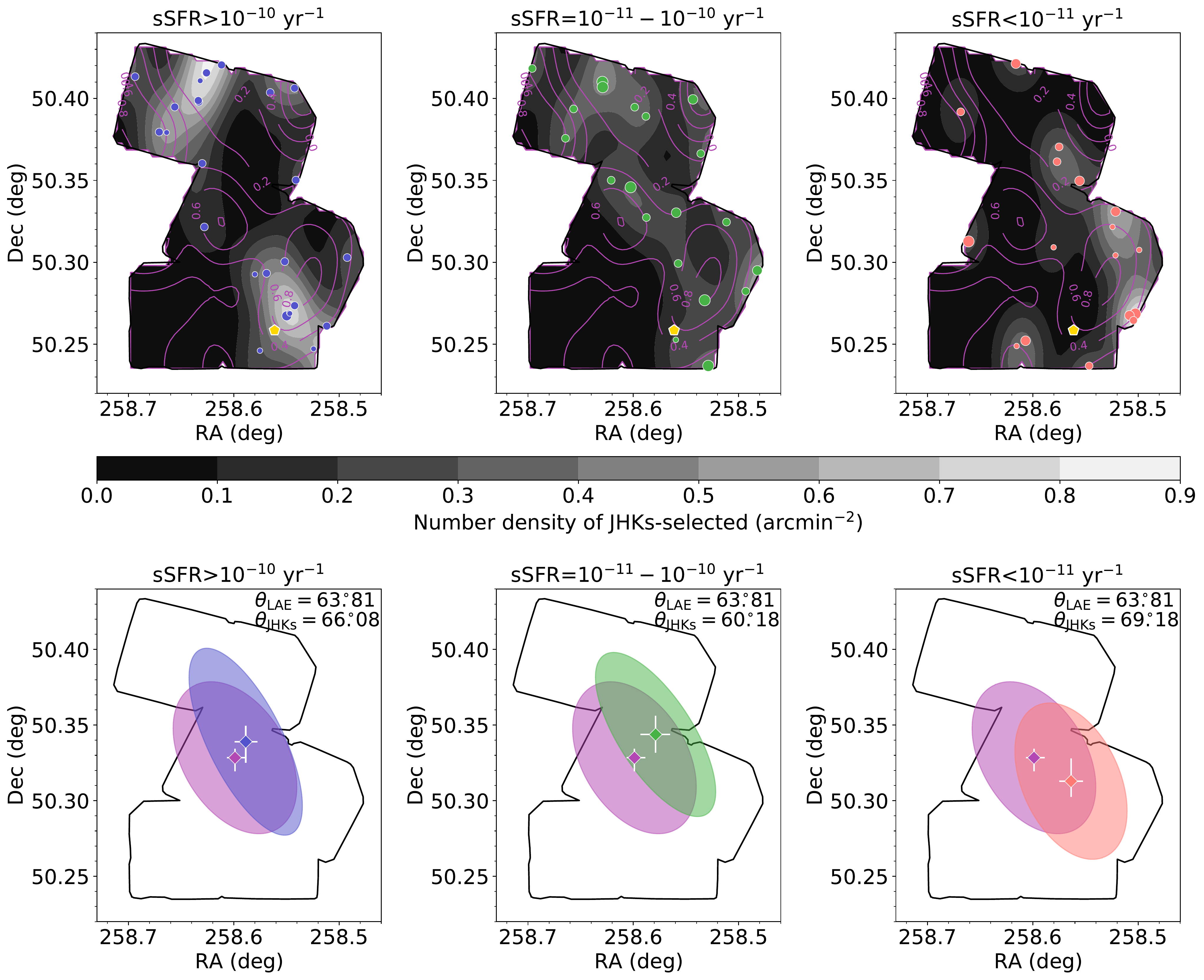}
  \caption{Upper panels: The same as left panel of Figure \ref{fig:fig11}, but the $JHK_s$-selected galaxies with different sSFRs are shown separately in the different panels (Upper left: sSFR $>10^{-10}\ \rm{yr^{-1}}$. Upper middle: sSFR $=10^{-11}$--$10^{-10}\ \rm{yr^{-1}}$. Upper right: sSFR $<10^{-11}\ \rm{yr^{-1}}$).
  Lower panels: The same as right panel of Figure \ref{fig:fig11}, but for the selected galaxies with different sSFRs.}
  \label{fig:fig12}
\end{figure*}

Panels (a)--(d) of Figure \ref{fig:fig9} show mass-weighted age vs.~$M_s$ and its probability density distribution.
The different colors indicate the $JHK_s$-selected galaxies with different sSFRs.
We also show the stellar mass completeness limit as a function of mass-weighted age.
More massive galaxies tend to have older stellar ages in the both fields.
In the 53W002 field, most of galaxies with $M_s>5\times10^{10}$ show ages older than $10^{8.5}\ \rm{yr}$, and six of the seven massive galaxies with $M_s>10^{11}\ M_\odot$ are older than 1 Gyr.
On the other hand, low-mass galaxies with $M_s<10^{10}\ M_\odot$ tend to show the mass-weighted ages younger than $10^{8.5}\ \rm{yr}$, although the scarcity of low-mass galaxies with old mass-weighted ages can be caused by the incompleteness.
Panels (e) and (f) of Figure \ref{fig:fig9} show the similar probability density excess map in the 53W002 field as the panel (e) of Figure \ref{fig:fig8} and the variation of probability density among the random fields of view in the COSMOS field, respectively.
As seen in Figure \ref{fig:fig8}, the distributions in the two fields are different at $M_s>10^{10.5}\ M_\odot$ and $<10^{10}\ M_\odot$.
In the excess at $M_s>10^{10.5}\ M_\odot$, the expected age increases with increasing stellar mass, and the probability is higher than the field-to-field variation, especially for galaxies with $M_s\sim7-10\times10^{10}\ M_\odot$.
At $M_s<10^{10}\ M_\odot$, the probability excess ranges from 5$\times10^{7}\ \rm{yr}$ to $5\times10^{8}\ \rm{yr}$.
Although there is a deficit of $\sim0.003$ arcmin$^{-2}$ at $M_s\sim2-5\times10^{10}\ M_\odot$ and mass-weighted age $>10^{9}$ yr, this deficit is smaller than the field-to-field variation.

\begin{deluxetable*}{cccccc}
  \tablecaption{Summary of the mean position and shape parameters.}\label{tab:table4}
  \tablewidth{0pt}
  \tablehead{
    \colhead{sample} &
    \colhead{$\Delta$RA\tablenotemark{a} (arcmin)} &
    \colhead{$\Delta$Dec\tablenotemark{a} (arcmin)} &
    \colhead{$A$ (arcmin)} & 
    \colhead{$B$ (arcmin)} &
    \colhead{$\theta$ (deg)}\smallskip
  }
  %\decimalcolnumbers
  \startdata
  LAEs & $0.00^{+0.29}_{-0.39}$ & $0.00^{+0.36}_{-0.54}$ & $3.22^{+1.34}_{-1.05}$ & $1.94^{+0.91}_{-0.60}$ & $63.81^{+5.56}_{-7.64}$\smallskip
  \\\hline
  $JHK_s$-selected galaxies & & & & & 
  \\
  all & $-0.78^{+0.24}_{-0.32}$ & $0.30^{+0.46}_{-0.49}$ & $3.79^{+1.27}_{-1.05}$ & $1.54^{+0.65}_{-0.53}$ & $64.76^{+3.66}_{-4.03}$\smallskip
  \\
  sSFR $>10^{-10}$ yr$^{-1}$ & $-0.39^{+0.40}_{-0.42}$ & $0.63^{+0.64}_{-0.83}$ & $4.01^{+1.48}_{-1.26}$ & $1.38^{+0.81}_{-0.63}$ & $66.08^{+3.95}_{-3.97}$\smallskip
  \\
  sSFR $=10^{-11}-10^{-10}$ yr$^{-1}$ & $-0.77^{+0.55}_{-0.52}$ & $0.92^{+0.75}_{-0.72}$ & $3.67^{+1.79}_{-1.43}$ & $1.39^{+0.81}_{-0.57}$ & $60.18^{+7.14}_{-5.95}$\smallskip
  \\
  sSFR $<10^{-11}$ yr$^{-1}$ & $-1.35^{+0.43}_{-0.44}$ & $-0.93^{+0.90}_{-0.62}$ & $3.24^{+1.87}_{-1.60}$ & $1.80^{+1.08}_{-0.65}$ & $69.18^{+6.36}_{-11.43}$\smallskip
  \enddata
  \tablecomments{The errors indicate the 68\% confidence range estimated with the bootstrap method.}
  \tablenotetext{a}{This is a coordinate originated at the mean position of LAEs.}
\end{deluxetable*}

\vspace{-8mm}
\subsection{Overlap with LAEs}\label{subsec:overlap_LAE}
Figure \ref{fig:fig10} shows the sky distribution of the $JHK_s$-selected galaxies and LAEs discovered by \citet{Mawatari_2012} in the 53W002 field.
The symbol size and color of circles represent the different $K_s$-band total magnitudes and $J-K_s$ colors of the selected objects, respectively.
The crosses and grey-scale contour show LAEs in the field of view of MOIRCS and their number density map.
The yellow pentagon indicates the radio galaxy 53W002.

The circle and cross symbols at the same position in the figure mean that the object satisfies both our $JHK_s$-bands color and $\Delta\chi^2_{z=2.39}$ criteria and the LAE selection ones.
Since our NIR selection basically picks up those galaxies with $M_s\gtrsim10^{10}\ M_\odot$ and is complementary to the LAE selection, the fraction of such overlapping objects is expected to be small.
Nevertheless, there are three $JHK_s$-selected galaxies that are also selected as LAEs.
Interestingly, all the three objects have bright $K_s$-band magnitudes of $K_s<22.25$.
One is the object No.18 in \citet{Pascarelle_1996b}, which is located about $40^{\prime\prime}$ northwest of 53W002 and one of the spectroscopically confirmed protocluster members at $z=2.393$.
Its rest-UV and optical emission lines indicate the AGN contribution \citep{Pascarelle_1996a,Motohara_2001}, and it has very extended Ly$\alpha$ emission \citep{Mawatari_2012}.
The estimated stellar mass is large ($9.4\times10^{10}\ M_\odot$), and its relatively blue color of $J-K_s\sim1.1$ may be affected by the AGN emission.
The second object is located about $50^{\prime\prime}$ northwest of the No.18.
This object has a red color of $J-K_s=2.2$ and a very large stellar mass of $2.1\times10^{11}\ M_\odot$.
Its estimated sSFR is intermediate ($10^{-10.6}\ \rm{yr^{-1}}$) with a moderate extinction of $E(B-V)=0.18$.
These properties suggest that its Ly$\alpha$ emission may be also related with AGN, although we need spectroscopy to confirm it.
The third object is located near the northwest corner of FOV3.
It has a stellar mass of $M_s=6.5\times10^{10}\ M_\odot$ and a relatively low sSFR of $10^{-10.8}\ \rm{yr^{-1}}$.
Interestingly, a Lyman $\alpha$ blob discovered by \citet{Mawatari_2012} is located within $\sim10^{\prime\prime}$ from this object.
These three objects have large stellar mass and old stellar age, which is different from typical LAEs.

We note that another faint $JHK_s$-selected galaxy with $K_s=22.9$ lies at $1\farcs5$ from LAE No.19 in \citet{Pascarelle_1996b} that is located about $10^{\prime\prime}$ northwest of the No.18.
The No.19 is QSO at $z=2.39$ \citep{Pascarelle_1996a}, and has a bright corresponding source in the $K_s$ band ($K_s$=20.3), which does not satisfy the $JHK_s$-color criteria probably due to the strong AGN emission.
The extended Ly$\alpha$ emission of the No.19 \citep{Mawatari_2012} seems to cover this object, and this may be a companion of the No.19.

\subsection{Spatial Distribution}
We here investigate the spatial distribution of the $JHK_s$-selected galaxies and compare  them with LAEs discovered by \citet{Mawatari_2012}.
In Figure \ref{fig:fig10}, the $JHK_s$-selected galaxies are distributed over our survey field except for southeast region (eastern half of FOV1), where the number density of the LAEs is also low.
The radio galaxy 53W002 is located at outskirt of the density peak of LAEs, and there are few bright and red galaxies in vicinity of 53W002, which is consistent with the previous study of \citet{Yamada_2001}.

In the left panel of Figure \ref{fig:fig11}, we show the distribution of the $JHK_s$-selected galaxies and the grey scale map of their number density.
While their distribution is widespread, there are two high-density regions of those galaxies.
One is located at southwest region of our survey field, namely, RA=$258.\!\!^{\circ}52$, Dec=$50.\!\!^{\circ}27$.
The other is located at northern region of the survey field, RA=$258.\!\!^{\circ}62$, Dec=$50.\!\!^{\circ}42$.
Interestingly, these two density peaks deviate from the two density peaks of LAEs in our survey field by $1.5$--$2.5$ arcmin ($\sim$0.7--1.2 physical Mpc at $z=2.39$), although the southwest high-density regions of LAEs and $JHK_s$-selected galaxies heavily overlap with each other.
The structures of the $JHK_s$-selected galaxies in and around the both high-density regions extend toward these density peaks of LAEs.

In order to characterize their spatial distributions, we calculated the mean positions and shape parameters based on the second-order moments for the $JHK_s$-selected galaxies and LAEs in our survey field.
We represent a spatial distribution as an ellipse with the shape parameters, namely, semi-major axis $A$, semi-minor axis $B$, and position angle between $A$ and the horizontal axis $\theta$.
We set  $x$ and $y$ coordinates where $x$ and $y$ measure a distance from the origin of RA=$258.\!\!^{\circ}599$, Dec=$50.\!\!^{\circ}328$, which corresponds to the mean position of LAEs, in the RA and Dec direction, respectively, and computed the shape parameters as follows:
\begin{eqnarray}
  \tan2\theta &=& \frac{2\overline{xy}}{\overline{x^2}-\overline{y^2}}
  \label{eq:eq_theta}
  \\
  A^2 = \frac{\overline{x^2}+\overline{y^2}}{2} &+& \sqrt{\left( \frac{\overline{x^2}-\overline{y^2}}{2} \right)^2 + \overline{xy}^2}
  \label{eq:eq_A2}
  \\
  B^2 = \frac{\overline{x^2}+\overline{y^2}}{2} &-& \sqrt{\left( \frac{\overline{x^2}-\overline{y^2}}{2} \right)^2 + \overline{xy}^2},
  \label{eq:eq_B2}
\end{eqnarray}
where $\overline{x^2}$, $\overline{{y^2}}$, and $\overline{xy}$ are the second-order moments of sample, respectively.
These second-order moments are calculated as follows:
\begin{eqnarray}
  \overline{x^2} &=& \frac{\sum_{i=1}^{N}(x_i-\overline{x})^2}{N}
  \label{eq:eq_x2}
  \\
  \overline{y^2} &=& \frac{\sum_{i=1}^{N}(y_i-\overline{y})^2}{N}
  \label{eq:eq_y2}
  \\
  \overline{xy} &=& \frac{\sum_{i=1}^{N}(x_i-\overline{x})(y_i-\overline{y})}{N},
  \label{eq:eq_xy}
\end{eqnarray}
where $x_i$, $y_i$, and $N$ are $x$ and $y$ values of each object and number of objects in the sample, respectively.
The right panel of Figure \ref{fig:fig11} shows the mean positions and ellipses with estimated shape parameters for the $JHK_s$-selected galaxies and LAEs in the field, and these parameters and their errors are summarized in Table \ref{tab:table4}.
The mean position of the $JHK_s$-selected galaxies deviates slightly but significantly from that of LAEs, while the separation is smaller than the semi-minor axis of the both distributions and they heavily overlap with each other.
The position angles of the both distributions are similar and consistent within the uncertainty.
The $JHK_s$-selected galaxies show a marginally larger semi-major axis (and smaller semi-minor axis), which probably reflects that the distance between the two density peaks mentioned above is larger for the $JHK_s$-selected galaxies; the northeast peak of the $JHK_s$-selected galaxies is located north or northwest of the peak of LAEs, while the southeast one is located south or southwest of that of LAEs.

We also investigated how the spatial distribution of the $JHK_s$-selected galaxies depends on their sSFR.
Figure \ref{fig:fig12} shows the distributions, mean positions, and ellipses with the estimated shape parameters of those galaxies with different sSFRs, separately.
The distribution of those with sSFR $>10^{-10}\ \rm{yr^{-1}}$ shows two clear high-density structures which extend to the density peaks of LAEs mentioned above.
Most of these star-forming galaxies seem to belong to these structures.
There are also two galaxies in the northwest part of FOV3 where the LAE density increases, although we cannot confirm how many these galaxies are located in this high-density region of LAEs due to our limited survey area.
The mean position and position angle of these star-forming galaxies are similar to those of LAEs, and most of their distribution overlap with that of LAEs although their semi-major axis is marginally larger than that of LAEs.
On the other hand, quiescent galaxies with sSFR $<10^{-11}\ \rm{yr^{-1}}$ tend to avoid the high-density region of LAEs.
Many such galaxies are distributed west and north of the southwest high-density peak of LAEs.
The mean position of these quiescent galaxies is located $1\farcm6$ southwest of that of LAEs.
The separation between the mean positions is comparable to semi-minor axis of the both distributions, and the displacement between the two distributions is significant.
The distribution of the $JHK_s$-selected galaxies with sSFR $=10^{-11}$--$10^{-10}\ \rm{yr^{-1}}$ overlaps with the both star-forming and quiescent galaxies mentioned above.
The mean position is located $1\farcm2$ northwest of that of LAEs, while the shape of their distribution is similar with those with sSFR $>10^{-10}$ yr$^{-1}$.

\begin{figure}[ht!]
  \centering
  \includegraphics[width=8.7cm]{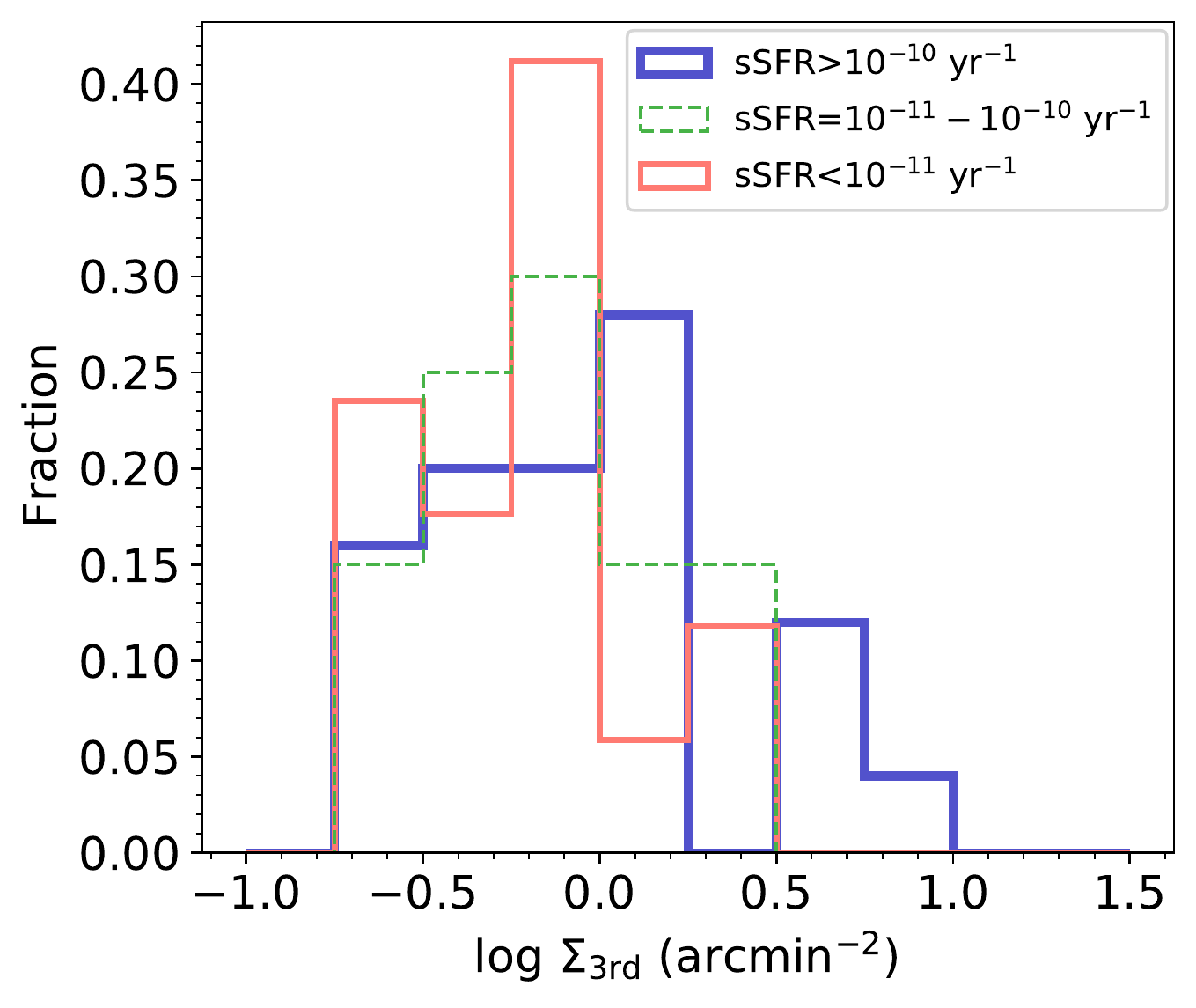}
  \caption{The distributions of the local number density of LAEs, $\Sigma_{3\rm{rd}}$ for the $JHK_s$-selected galaxies with different sSFRs. The blue, green, and red histograms mean the distributions for those with sSFR $>10^{-10}$ yr$^{-1}$, sSFR $=10^{-11}-10^{-10}$ yr$^{-1}$, and sSFR $<10^{-11}$ yr$^{-1}$, respectively.}
  \label{fig:fig13}
\end{figure}
We also estimated local number densities of LAEs at the positions of the $JHK_s$-selected galaxies to investigate their relationship with LAEs more locally.
We measured the local number density of LAEs by searching for three nearest LAEs and calculating surface number density within a radius of the distance to the third nearest neighbor as follows:
\begin{eqnarray}
  \Sigma_{3\rm{rd}} = \frac{3}{\pi r_{3\rm{rd}}^2},
  \label{eq:eq_nsigma}
\end{eqnarray}
where $r_{3\rm{rd}}$ is the distance from the $JHK_s$-selected galaxy to the third nearest LAE.
In cases where the $JHK_s$-selected galaxy satisfies the LAE selection, we did not count the galaxy as a LAE, because AGN could mainly contribute to the Ly$\alpha$ emission of these objects (Section \ref{subsec:overlap_LAE}).
Figure \ref{fig:fig13} shows the distributions of $\Sigma_{3\rm{rd}}$ for the $JHK_s$-selected galaxies with different sSFRs.
One can see that star-forming galaxies with high sSFRs show higher fraction of those with high $\Sigma_{\rm{3rd}}$ than those with lower sSFRs.
The fraction of those with $\log\Sigma_{\rm{3rd}}>0$ is 44\% for star-forming galaxies with sSFR $>10^{-10}$ yr$^{-1}$.
On the other hand, 30\% and 18\% of those galaxies with sSFR $=10^{-11}$--$10^{-10}$ yr$^{-1}$ and sSFR $<10^{-11}$ yr$^{-1}$ show $\log\Sigma_{\rm{3rd}}>0$, respectively.
Those quiescent galaxies with low sSFRs tend to have a relatively large separation from LAEs, while active star-forming galaxies show various local densities of LAEs.

%%%%%%%%%%%%%%%%%%%%%%%%%%%%%%%%%%%%%%%%%%%%%%%%%%%%%%%%%%%%%%%%%%
%%%%%%%                                                    %%%%%%%
%%%%%%%                                                    %%%%%%%
%%%%%%%                     Discussion                     %%%%%%%
%%%%%%%                                                    %%%%%%%
%%%%%%%                                                    %%%%%%%
%%%%%%%%%%%%%%%%%%%%%%%%%%%%%%%%%%%%%%%%%%%%%%%%%%%%%%%%%%%%%%%%%%
\section{Discussion} \label{sec:discussion}

\subsection{Bright and Red Galaxies}
\begin{figure}[ht!]
  \centering
  \includegraphics[width=8cm]{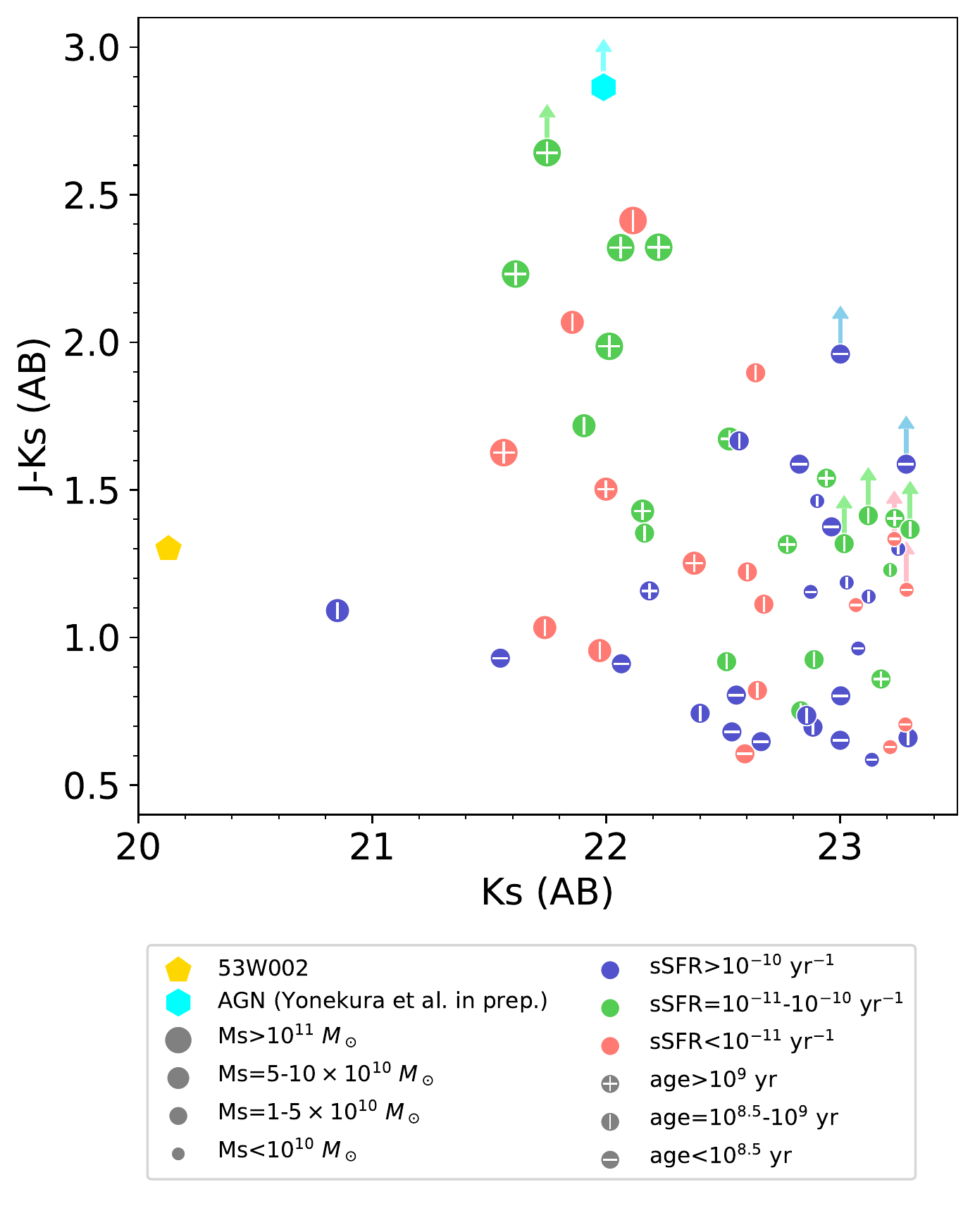}
  \caption{$J-K_s$ vs.~$K_s$ color magnitude diagram for the $JHK_s$-selected galaxies in the 53W002 field.
  The different colors and sizes of symbols represent their sSFRs and stellar mass, while cross, vertical, and horizontal lines mean the different mass-weighted ages.
  The yellow pentagon and cyan hexagon indicate the radio galaxy 53W002 and AGN (see text), respectively.}
  \label{fig:fig14}
\end{figure}
\begin{figure}[ht!]
  \centering
  \includegraphics[width=8.7cm]{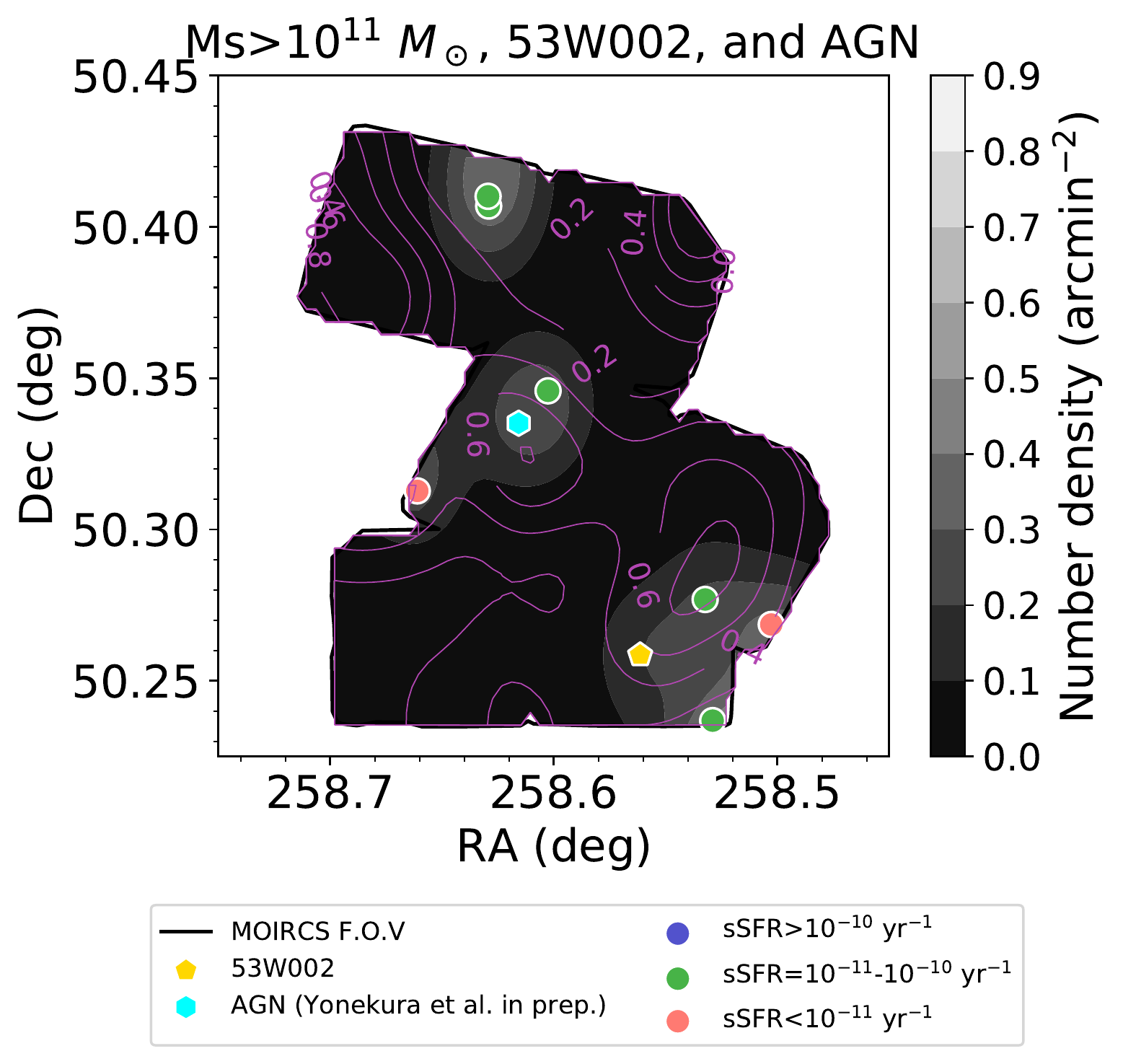}
  \caption{The same as the left panel of Figure \ref{fig:fig11}, but the selected galaxies with $M_s>10^{11}\ M_\odot$ and other massive galaxies are shown. The yellow pentagon and cyan hexagon indicate the radio galaxy 53W002 and AGN (see text), respectively. The grey contour shows the number density of these galaxies including the 53W002 and AGN.}
  \label{fig:fig15}
\end{figure}
We found that the $JHK_s$-selected galaxies with bright $K_s$-band magnitudes and/or red $J-K_s$/$V-K_s$ colors show the number density excesses in the 53W002 field (Figure \ref{fig:fig5}).
In particular, those galaxies with $K_s<22.25$ and $J-K_s>2$ show about nine times higher number density than in the COSMOS field (Table \ref{tab:table3}).
In Figure \ref{fig:fig14}, we plot the selected galaxies with different $M_s$, sSFRs, and mass-weighted ages by different symbols on the $J-K_s$ vs.~$K_s$ plane.
Five out of the six galaxies with $K_s<22.25$ and $J-K_s>2$ have $M_s>10^{11}\ M_\odot$, and the stellar mass of the other one is $8\times10^{10}\ M_\odot$.
There are additional two massive galaxies with $M_s>10^{11}\ M_\odot$, which do not satisfy $J-K_s>2$, but have similarly bright magnitudes and red colors.
Most of these galaxies show old mass-weighted stellar ages of 1.5--2.0 Gyr.
On the other hand, their sSFRs are intermediate values of $10^{-11}$--$10^{-10}\ \rm{yr^{-1}}$, which suggests that their star formation has not completely stopped.
Their best-fit dust extinction values are $E(B-V)=0.18$--$0.42$, and their very red colors of $J-K_s>2$ are caused by combination of the old stellar population and dust extinction.
We note that there are two more protocluster members with similarly large stellar mass of $\gtrsim10^{11}\ M_\odot$.
One is the radio galaxy 53W002 whose estimated stellar mass is $\sim1.8\times10^{11}\ M_\odot$, although the strong emission from AGN could affect our estimation with the SED fitting.
The other object was selected by the $JHK_s$-color criteria, but was not included in our sample because its $\Delta\chi^2_{z=2.39}$ is large probably due to strong emission lines.
This object is also one of LAEs selected by \citet{Mawatari_2012}, and we confirmed very strong [OIII]$\lambda\lambda4959,5007$ emission lines at $z=2.39$ in its NIR spectrum, which suggests this object also has AGN (Yonekura et al. in prep.).

Many massive galaxies having $M_s>10^{11}\ M_\odot$ are associated with the protocluster with a high number density, which cannot be seen in the COSMOS field.
Figure \ref{fig:fig15} shows their spatial distribution.
Their spatial distribution does not concentrate on some position in the protocluster, but they are rather widely distributed along a SW-NE direction.
While the sky positions of these massive galaxies tend to deviate from the density peaks of LAEs, they are located along the similar direction with the structure of LAEs.
These massive galaxies have old stellar ages but have not yet stopped their star formation completely, and some of these have AGN activity as mentioned above.
On the other hand, most of LAEs are expected to be young and formed very recently.
In this structure along the SW-NE direction, there might be large-scale gas supply, which leads to the coherent formation of LAEs and maintenance of the activities in massive galaxies (Section \ref{subsec:5p4}).

If this protocluster evolves into a massive cluster at lower redshift, these massive galaxies with $M_s>10^{11}\ M_\odot$ are expected to form the bright end of the red sequence.
At $z=2.39$, they show a large scatter in $J-K_s$ color probably due to their star formation and dust extinction in spite of their old mean age, and their $J-K_s$ colors seem to be too red for the red sequence at $z\sim2.4$ \citep{Kodama_2007}.
In fact, (slightly less massive) quiescent galaxies with similarly old ages show $J-K_s\sim1.5$ (Figure \ref{fig:fig14}).
If the star formation in the massive galaxies would stop as this structure grows and collapses, these massive galaxies with very red colors will lose gas and dust and concentrate on the red sequence.

\subsection{Quiescent Galaxies}
In addition to the most massive galaxies with $M_s>10^{11}\ M_\odot$, we also found the density excess of relatively massive quiescent galaxies with $M_s=5\times10^{10}$--$10^{11}\ M_\odot$ and sSFR $<10^{-11}\ \rm{yr^{-1}}$ in the 53W002 field (panel (e) of Figure \ref{fig:fig8}).
Their mass-weighted stellar ages are $\sim0.5$--$2.0$ Gyr.
They show a variety of $J-K_s$ color depending on their stellar age and dust extinction (Figure \ref{fig:fig14}), while those with old stellar ages of $\gtrsim1$ Gyr tend to have $J-K_s\sim1.5$ as mentioned above.
The sSFRs of these quiescent galaxies are systematically lower than those of field galaxies with similar stellar masses of $5\times10^{10}$--$10^{11}\ M_\odot$ in the COSMOS field (panels (c) and (d) of Figure \ref{fig:fig8}).
While we fixed the redshift to $z=2.39$ in the SED fitting for these field galaxies to subtract their contribution from the results in the 53W002 field, the estimated $M_s$, sSFR, and mass-weighted age do not significantly change if we adopt the photometric redshifts from the COSMOS2015 catalog in the SED fitting and select only those at $z_{photo}=2.39\pm0.20$ (Appendix \ref{appendix:fixed_redshift}).
On the other hand, the mass-weighted ages of these quiescent galaxies in the 53W002 field are similar with those of field galaxies with similar mass.
Therefore, these quiescent galaxies may have experienced similar star formation histories with field galaxies in the past and then stopped their star formation recently, for example, hundreds Myr before the observed epoch.
These quiescent galaxies tend to avoid the high-density region of LAEs, and many of them are located north and west of southwest density peak of LAEs (upper right panel of Figure \ref{fig:fig12}).
Interestingly, most of low-mass quiescent galaxies with $M_s<10^{10}\ M_\odot$ and sSFR $<10^{-11}\ \rm{yr^{-1}}$ are also located in the same region.
They show a young stellar age of $\lesssim10^{8.5}$ yr and $E(B-V)\sim0$, which is similar with low-mass star-forming galaxies with $M_s<10^{10}\ M_\odot$ and sSFR $>10^{-10}\ \rm{yr^{-1}}$.
Therefore, their lower sSFRs are probably caused by recent quenching of star formation.
While quiescent galaxies with $M_s\lesssim10^{11}\ M_\odot$ are located in this region, we do not know how far the region of quiescent galaxies extends due to our limited survey area.
There may be a massive structure or main halo west of our survey area, where some mechanisms such as starburst by merger/interaction, starvation, or tidal/ram-pressure stripping might work well \citep[e.g.,][]{Muldrew_2018}.

\subsection{Low-Mass Star-Forming Galaxies}
The low-mass galaxies with $M_s<10^{10}\ M_{\odot}$ in the 53W002 field also show a number density excess.
The number density of these low-mass galaxies is more than three times higher than that of those galaxies in the COSMOS field (Figure \ref{fig:fig7}).
The number density excess of low-mass galaxies can be seen over a wide range of sSFR (Figures \ref{fig:fig7} and \ref{fig:fig8}). 
Although the excess of low-mass galaxies with sSFR $>10^{-10}\ \rm{yr^{-1}}$ is less significant than those with sSFR $<10^{-10}\ \rm{yr^{-1}}$, many of these low-mass star-forming galaxies are located near or within the high-density regions of LAEs (Figure \ref{fig:fig12}), which suggests that they are associated with this protocluster.
These star-forming galaxies have young mass-weighted ages of $\lesssim10^{8.5}\ \rm{yr}$ and very small dust extinction (median value of $E(B-V)=0.02$).
These properties are similar with LAEs, while most LAEs are expected to have much smaller mass and younger age \citep[e.g.,][]{Overzier_2008,Ono_2010,Hagen_2014,Shimakawa_2017a}.
Such similarity in the physical properties may be related with the heavily overlapping spatial distributions of these galaxies and LAEs.

We note that our NIR data are heavily incomplete for low-mass galaxies with an old stellar age of $\gtrsim10^9$ yr (Figure \ref{fig:fig9}).
Deeper optical and NIR data are needed to probe such low-mass galaxies with old stellar population.

\subsection{Intermediate-Mass Galaxies}
While massive and low-mass galaxies show the number density excesses in the 53W002 field as described above, the number density of the $JHK_s$-selected galaxies with $1$--$5\times10^{10}\ M_\odot$ in the 53W002 field is similar with that in the COSMOS field showing no significant excess.
Since the field-to-field variation is relatively large in this stellar mass range for our selection method (panel (f) of Figure \ref{fig:fig8}), the relatively lower density of these galaxies could be explained by the field variation of foreground/background galaxies.
Otherwise, the deficit of these intermediate-mass galaxies may occur in this protocluster.
\citet{Shi_2021} reported the similar bimodal number density excess in the stellar mass function and deficit between the massive and low-mass excesses in a protocluster originally selected by a number density excess of LBGs at $z=3.24$, although the deficit of galaxies occurs at more massive stellar mass of $10^{10.6}$--$10^{11}\ M_\odot$.
If this is the case in the 53W002 protocluster, some mechanism decreasing their number density such as active mergers and/or disruption in earlier time \citep[e.g.,][]{Muldrew_2018} may be required because the stellar mass function of galaxies at similar redshifts in general fields does not show such deficit \citep[e.g.,][]{Kajisawa_2009,Ilbert_2013}.
We also note that some fraction of these intermediate-mass galaxies could be missed in this study, if their mass-weighted ages are very old, for example, $\gtrsim$ 1 Gyr (Figure \ref{fig:fig9}).

\subsection{Comparisons with Previous Studies}\label{subsec:5p4}
We searched for massive galaxies with the optical-NIR photometric data in the field of the high-density structure of LAEs at $z=2.39$ near the radio galaxy 53W002, and found the significant number density excess of massive galaxies with $M_s>10^{11}\ M_\odot$ as well as quiescent galaxies with $M_s=5\times10^{10}$--$10^{11}\ M_\odot$ and low-mass galaxies with $M_s<10^{10}\ M_\odot$.
Our result indicates a possibility that such high-density structures of LAEs, which tend to be low-mass and young galaxies, can trace protoclusters which include more massive galaxies.
Previously, \citet{Uchimoto_2012} and \citet{Kubo_2013} carried out the similar search for massive galaxies in the high-density region of LAEs at $z=3.09$ in the SSA22 field \citep{Hayashino_2004,Yamada_2012}, and found a significant number density excess of massive galaxies along the structure of LAEs, which includes a concentration of quiescent galaxies near the density peak of LAEs.
The structure of LAEs in the SSA22 field has large extent ($\sim$ 60 comoving Mpc) and very high-density peak \citep{Yamada_2012}.
\citet{Mawatari_2012} reported that the overdensity of LAEs in the SSA22 protocluster corresponds to a rareness probability of 0.0017\%, while that of LAEs in the 53W002 field is 0.9\%.
\citet{Shi_2019a} also searched for massive galaxies in a rich protocluster of LAEs at $z=3.78$, which was originally discovered as a concentration of five bright LBGs at $z=3.78$ \citep{Lee_2013,Lee_2014,Dey_2016}, and found a significant high-density structure of photo-$z$ selected star-forming galaxies and a number density excess of color-selected Balmer break galaxies with $M_s>10^{11}\ M_\odot$.
Our discovery of the number density excess of massive galaxies in the 53W002 field suggests that more moderate high-density structures of LAEs could also trace protoclusters with many massive galaxies.

The similar searches for massive galaxies in protoclusters of LAEs have also been done in the fields of high-redshift powerful radio galaxies at $z\sim2$--$3$ \citep{Kurk_2004,Kodama_2007}.
These studies found number density excesses of NIR-selected (massive) galaxies in the protoclusters of LAEs, which had been discovered by the narrow-band search around the radio galaxies \citep{Venemans_2007}.
\citet{Kodama_2007} found that massive galaxies tend to be located along the structure of LAEs.
We note that such powerful radio galaxies at high redshift are preferentially located in high-density environments \citep[e.g,][]{Hatch_2014}, and therefore the overdensity of massive galaxies in these studies may be closely related with the radio galaxies rather than the structure of LAEs.
While massive galaxies and LAEs in these protoclusters are distributed around the powerful radio galaxies, the radio galaxy 53W002 is located at outskirt of the high-density structures of both LAEs and $JHK_s$-selected galaxies.
The distribution of these massive galaxies in the 53W002 field seems to be related with the structures of LAEs rather than the radio galaxy.

Several previous studies reported the spatial segregation between LAEs and more massive galaxies in protoclusters at the similar redshifts \citep[e.g.,][]{Yang_2010,Shimakawa_2017b,Shi_2019a,Shi_2020}.
\citet{Shimakawa_2017b} reported that LAEs tend to avoid the highest-density regions of more massive H$\alpha$ emitters in a protocluster, which consists of large number of star-forming galaxies at $z=2.53$ \citep{Hayashi_2012}.
\citet{Shi_2019a} found that a peak position of the overdensity of photo-$z$ selected (relatively massive) star-forming galaxies deviates from the high-density region of LAEs, and color-selected quiescent galaxies also avoid the overdensity of LAEs.
\citet{Shi_2020} reported that photo-$z$ selected (massive) quiescent galaxies are segregated from the high-density region of LAEs, while less-dusty star-forming galaxies with relatively low stellar mass coexist with LAEs in the high-density region.
In the 53W002 field, the distribution of star-forming galaxies with sSFR $>10^{-10}$ yr$^{-1}$ heavily overlaps with the structure of LAEs, but the density peaks of these galaxies slightly deviate from the peaks of LAEs, which is similar with the results of previous studies.
Furthermore, the quiescent galaxies with sSFR $<10^{-11}\ \rm{yr^{-1}}$, which include low-mass galaxies with $M_s<10^{10}\ M_\odot$, are located avoiding the structure of LAEs (Figure \ref{fig:fig12} and \ref{fig:fig13}).
Such segregation may be caused by different evolutionary phases/mass assembly histories of different dark matter halos in the protocluster \citep[e.g.,][]{Muldrew_2018,Shi_2019b}.
For example, \citet{Muldrew_2018} studied the evolution of galaxies in protoclusters with their semi-analytic model, and proposed that the quenching of low-mass galaxies could occur mainly in a main (the most massive) halo in protoclusters at $z>2$.
The region dominated by quiescent galaxies seen in Figure \ref{fig:fig12} may be associated with such a massive halo in the protocluster.

Since LAEs are generally very young as mentioned above, those LAEs in the high-density structure are expected to almost simultaneously form very recently.
Such synchronized formation of galaxies may need a recent cold gas supply over a scale of the structure.
On the other hand, the H{\footnotesize I} and dust column densities should be sufficiently low so that we can observe their Ly$\alpha$ emission and identify the structure of LAEs \citep[e.g.,][]{Shimakawa_2017b}.
In fact, several studies reported that the large-scale (tens of Mpc) H{\footnotesize I} gas traced by Ly$\alpha$ absorption is associated with high-density structures of LAEs \citep[][]{Cucciati_2014,Mawatari_2017,Liang_2021}, while the high optical depth of Ly$\alpha$ emission could lead to a deficit of LAEs at small scale \citep[e.g.,][]{Momose_2021a,Momose_2021b}.
The most massive galaxies with $M_s>10^{11}\ M_\odot$ in the 53W002 field tend to avoid high-density peaks of LAEs, but they show the similar wide distribution along the SW-NE direction (Figure \ref{fig:fig15}).
While they have old mass-weighted stellar ages of 1.5--2.0 Gyr, their intermediate sSFRs
of $10^{-11}$--$10^{-10}\ \rm{yr^{-1}}$ suggest that their star-formation activities have not yet stopped.
In contrast, several studies reported that such massive protocluster members tend to be quiescent galaxies with lower sSFRs \citep[e.g.,][]{Tanaka_2010,Kubo_2013,Kubo_2015,Kubo_2021,Ando_2020,Chartab_2020,Shi_2021}.
Such star formation in the massive galaxies in the 53W002 field may be related with the large-scale gas supply along the structure of LAEs, although we don't know separations between massive galaxies and the structure of LAEs in the line of sight direction without spectroscopic redshifts.

The number density excess of the $JHK_s$-selected galaxies in the 53W002 field is significant at high stellar mass of $\gtrsim10^{11}\ M_\odot$, while low-mass galaxies with $M_s\lesssim10^{10}\ M_\odot$ also show the excess.
Several previous studies found that the stellar mass distribution of galaxies in protoclusters is skewed toward higher stellar mass (more top heavy) than field galaxies at the same redshifts \citep[e.g.,][]{Cooke_2014,Steidel_2005,Hatch_2011b,Shimakawa_2018,Ando_2020,Koyama_2021}.
Theoretical models also predict that high abundance of massive dark matter halos in high-density regions such as protoclusters promotes the early and active formation of massive galaxies, because the sufficiently high density in such halos enables to form stars even in early epoch (the biased galaxy formation) and the high merger rate among halos in such regions drives stellar mass growth of galaxies \citep{Lovell_2018,Muldrew_2018}.
The number density excess of massive galaxies in the 53W002 field seems to be consistent with such a scenario.
Their old ages (1.5--2.0 Gyr) and relatively low sSFRs ($10^{-11}$--$10^{-10}\ \rm{yr^{-1}}$) suggest that the formation of massive galaxies has been accelerated in this protocluster, while some previous studies discovered massive actively star-forming galaxies in other protoclusters \citep[e.g.,][]{Cooke_2014,Shimakawa_2018,Koyama_2021}.
Since these galaxies with $M_s>10^{11}\ M_\odot$ are distributed over a scale of $\sim$15--20 comoving Mpc, they are probably associated with different dark mater halos.
Many of them may be still central galaxies at the observed epoch, whose moderate (or declining) star formation activities may be maintained by the large-scale gas supply mentioned above.

%%%%%%%%%%%%%%%%%%%%%%%%%%%%%%%%%%%%%%%%%%%%%%%%%%%%%%%%%%%%%%%%%%
%%%%%%%                                                    %%%%%%%
%%%%%%%                                                    %%%%%%%
%%%%%%%                     Summary                        %%%%%%%
%%%%%%%                                                    %%%%%%%
%%%%%%%                                                    %%%%%%%
%%%%%%%%%%%%%%%%%%%%%%%%%%%%%%%%%%%%%%%%%%%%%%%%%%%%%%%%%%%%%%%%%%
\section{Summary} \label{sec:summary}
In this paper, we searched for massive galaxy candidates using the optical--NIR imaging data in the 53W002 protocluster field which had been discovered as the overdensity of LAEs at $z=2.39$.
We used the $JHK_s$-band color cuts and additional criterion from the SED fitting analysis to select galaxy candidates at $z=2.39$.
We also made the comparison sample in the general field from COSMOS2015 catalog with the same selection method and compared the statistical properties of the selected galaxies.
Our main results are as follows:

\begin{itemize}
  \item 62 and 896 galaxies with $K_s<23.3$ were selected in the 53W002 and COSMOS fields, respectively.
  The number density of the $JHK_s$-selected galaxies in the 53W002 field is significantly higher than that in the COSMOS field at $K_s<22.25$, $J-K_s>2$, or $V-K_s>4$.
  In particular, the number density of those with $K_s<22.25$ and $J-K_s>2$ is nine times higher than that in the general field.

  \item Our SED fitting analysis suggests that the objects with $K_s<22.25$ and $J-K_s>2$ have stellar mass of $M_s>10^{11}\ M_\odot$.
  While their mass-weighted ages are $\sim$1.5--2.0 Gyr, they show moderate sSFRs of $10^{-11}$--$10^{-10}\ \rm{yr^{-1}}$ and dust extinction of $E(B-V)\sim0.2$--$0.4$.

  \item We also found the number density excesses for the quiescent galaxies with $M_s=5\times10^{10}$--$10^{11}\ M_\odot$ and sSFR $<10^{-11}\ \rm{yr^{-1}}$, and for low-mass galaxies with $5\times10^9$--$10^{10}\ M_\odot$ over a wide range of sSFR.
  
  \item The spatial distribution of the $JHK_s$-selected galaxies strongly depends on their $M_s$ and sSFR.
  The massive galaxies with $M_s>10^{11}\ M_\odot$ tend to avoid the density peaks of LAEs, but show the similar wide distribution along the SW-NE direction with the structure of LAEs.
  Quiescent galaxies with $M_s<10^{11}\ M_\odot$ and sSFR $<10^{-11}\ \rm{yr^{-1}}$ clearly avoid the high-density regions of LAEs, and many such galaxies are located in the western part of the survey area.
  On the other hand, the distribution of low-mass star-forming galaxies with sSFR $>10^{-10}\ \rm{yr^{-1}}$ heavily overlaps with the structure of LAEs.
\end{itemize}

While our results suggest that massive galaxies are abundant in the protocluster selected by the relatively moderate high-density structure of LAEs, we definitively need spectroscopic confirmation for the membership of those galaxies in this field.
Our limited survey area also prevents us from understanding the overall relationship between such massive galaxies and the structure of LAEs.
Furthermore, we here studied only one protocluster, and more investigations are needed for understanding the high-density structures of LAEs and formation of massive galaxies in protoclusters.

%% Putting eqnarrays or equations inside the mathletters environment groups
%% the enclosed equations by letter. For instance, the eqnarray below, instead
%% of being numbered, say, (4) and (5), would be numbered (4a) and (4b).
%% LaTeX the paper and look at the output to see the results.

%\section{Revision tracking and color highlighting} \label{sec:highlight}

%\section{Software and third party data repository citations} \label{sec:cite}

\begin{acknowledgments}
  We thank the anonymous referee for the valuable comments and suggestions.
  The imaging data used in this paper were taken with the Subaru Telescope, which is operated by National Astronomical Observatory of Japan (NAOJ).
  Data analysis in this paper was carried out on the Multi-wavelength Data Analyssi System operated by the Astronomy Data Center (ADC), NAOJ.
  This work was financially supported by JSPS (17K05386).
  K.~M.~is supported by JSPS KAKENHI grant No.20K14516.
\end{acknowledgments}

\appendix

\section{Effects of the Nebular Emission}\label{appendix:line_effect}

\subsection{Sample Selection}\label{appendix:selection}
In the SED fitting for the calculation of $\Delta\chi^2_{z=2.39}$, we chose not to include the nebular emission in the model templates (Section \ref{subsec:sample}).
We here check the effects of including the nebular emission on the sample selection.
We used the model spectra of PANHIT \citep{Mawatari_2020}, which consist of the stellar component constructed by the GALAXEV library and the nebular emission component by the \citet{Inoue_2011} model.
The nebular emission is calculated from the ionizing photon production rate of the stellar component under the assumption of the same metallicity with the stellar component, typical ranges of hydrogen number density and ionizing parameter, plane-parallel geometry, and the escape fraction of ionizing photon of $f_C$ \citep{Inoue_2011}.
For example, when $f_C$ is 0.0, all produced ionizing photons are used for ionization.
On the other hand, when $f_C$ is 1.0, no ionizing photons are used.
We assumed 4 kinds of $f_C$ of 0.0, 0.5, 0.9, and 1.0.
Other parameters are the same values described in Section \ref{subsec:sample}.
We carried out the SED fitting with the models described above and selected the sample by using the $JHK_s$-color cuts and $\Delta\chi^2_{z=2.39}$ criterion in both 53W002 and COSMOS fields.

\begin{figure}[ht!]
  \centering
  \includegraphics[width=7.5cm]{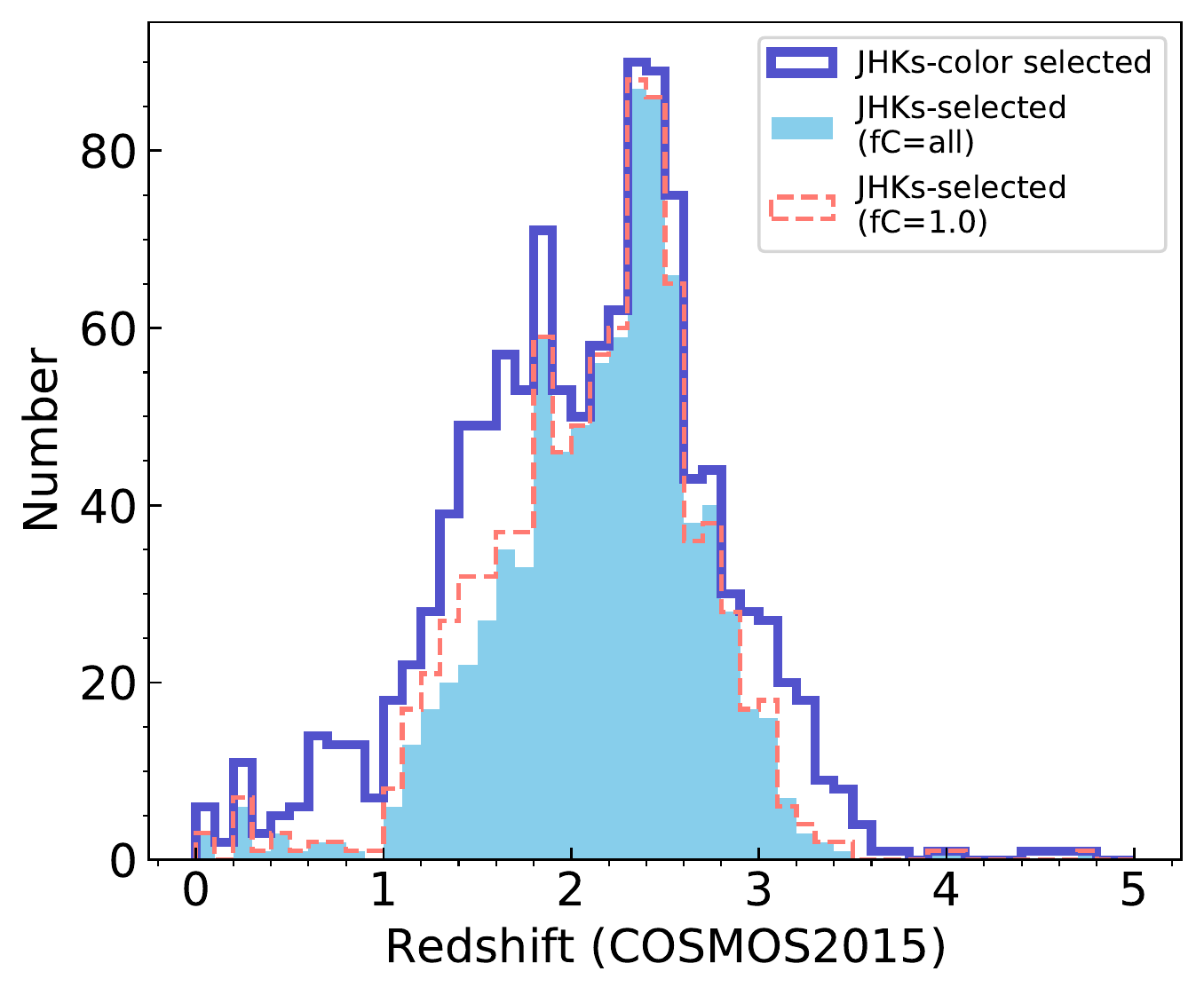}
  \caption{The redshift distribution of the selected objects in the COSMOS field. 
  The photometric redshifts from the COSMOS2015 catalog are used.
  The blue line histogram shows the distribution for galaxies selected by only the $JHK_s$-color cuts.
  Light blue shaded histogram shows those selected by the $JHK_s$-band color and $\Delta\chi^2$ criteria where the nebular emission is considered in the calculation of $\Delta\chi^2_{z=2.39}$. 
  Red dashed line histogram shows the distribution of galaxies selected by the original selection where the nebular emission is not included.}
  \label{fig:fig16}
\end{figure}

We selected 60 and 855 objects in the 53W002 and COSMOS fields, respectively.
Out of the selected galaxies, 59 and 805 objects are selected also by the original selection where the nebular emission was not included in the model templates.
Therefore, about 90\% of galaxies are selected regardless of the nebular emission in both fields.
Figure \ref{fig:fig16} shows the distributions of the photometric redshift from the COSMOS2015 catalog (hereafter $z_{\rm{COS}}$) for the selected galaxies in the COSMOS field.
Indeed, the redshift distribution doesn't change significantly if we consider the nebular emission.
While the number of contaminants from foreground/background redshifts is smaller when including the nebular emission, a small fraction of galaxies with $z_{\rm{COS}}\sim2.4$ are also missed.
\begin{figure*}[ht!]
  \centering
  \includegraphics[width=16cm]{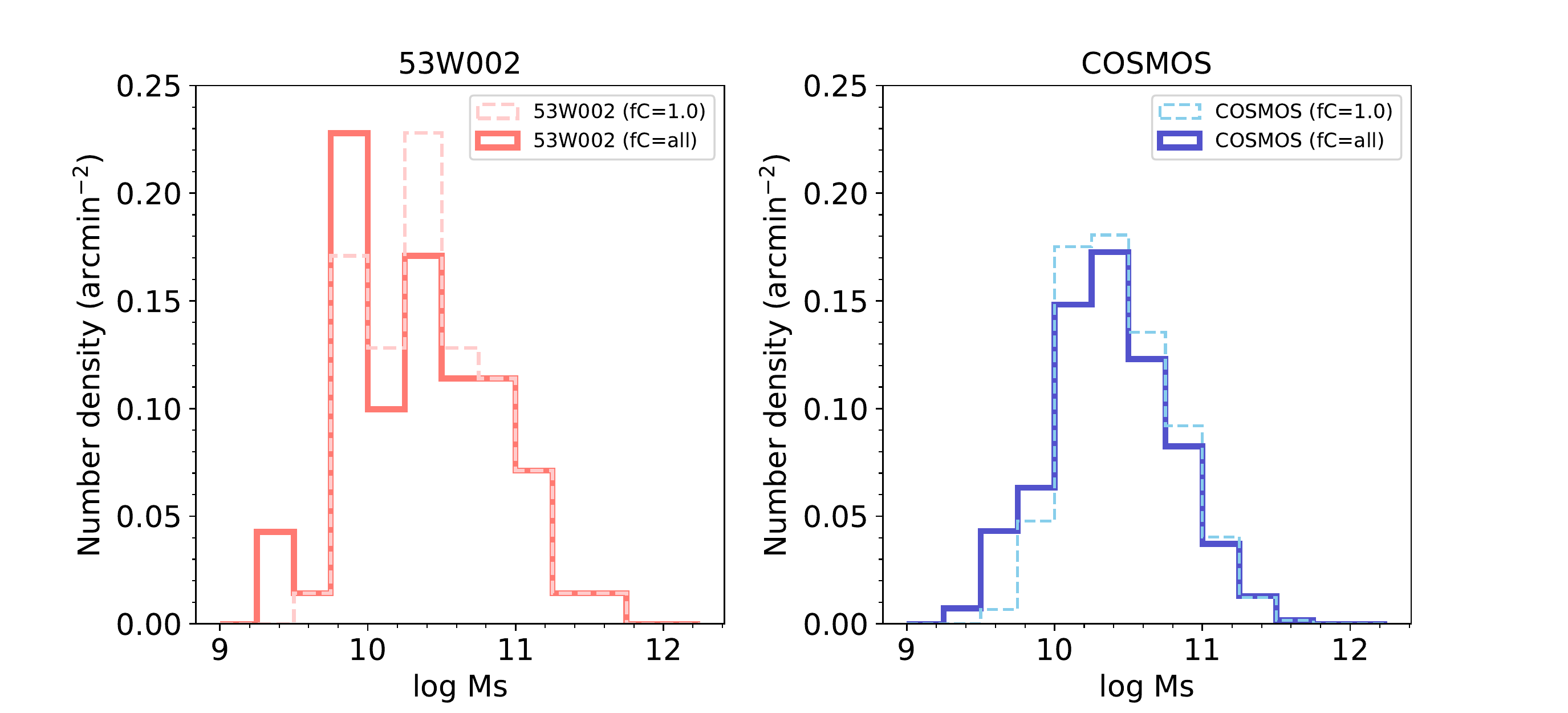}
  \caption{The stellar mass distributions of the $JHK_s$-selected galaxies in the cases with and without the nebular emission in the SED fitting.
  The thick solid histogram shows the case with the nebular emission, and the thin dashed histogram represents that without the nebular emission (original analysis).
  }
  \label{fig:fig17}
\end{figure*}
\begin{figure*}[ht!]
  \centering
  \includegraphics[width=17cm]{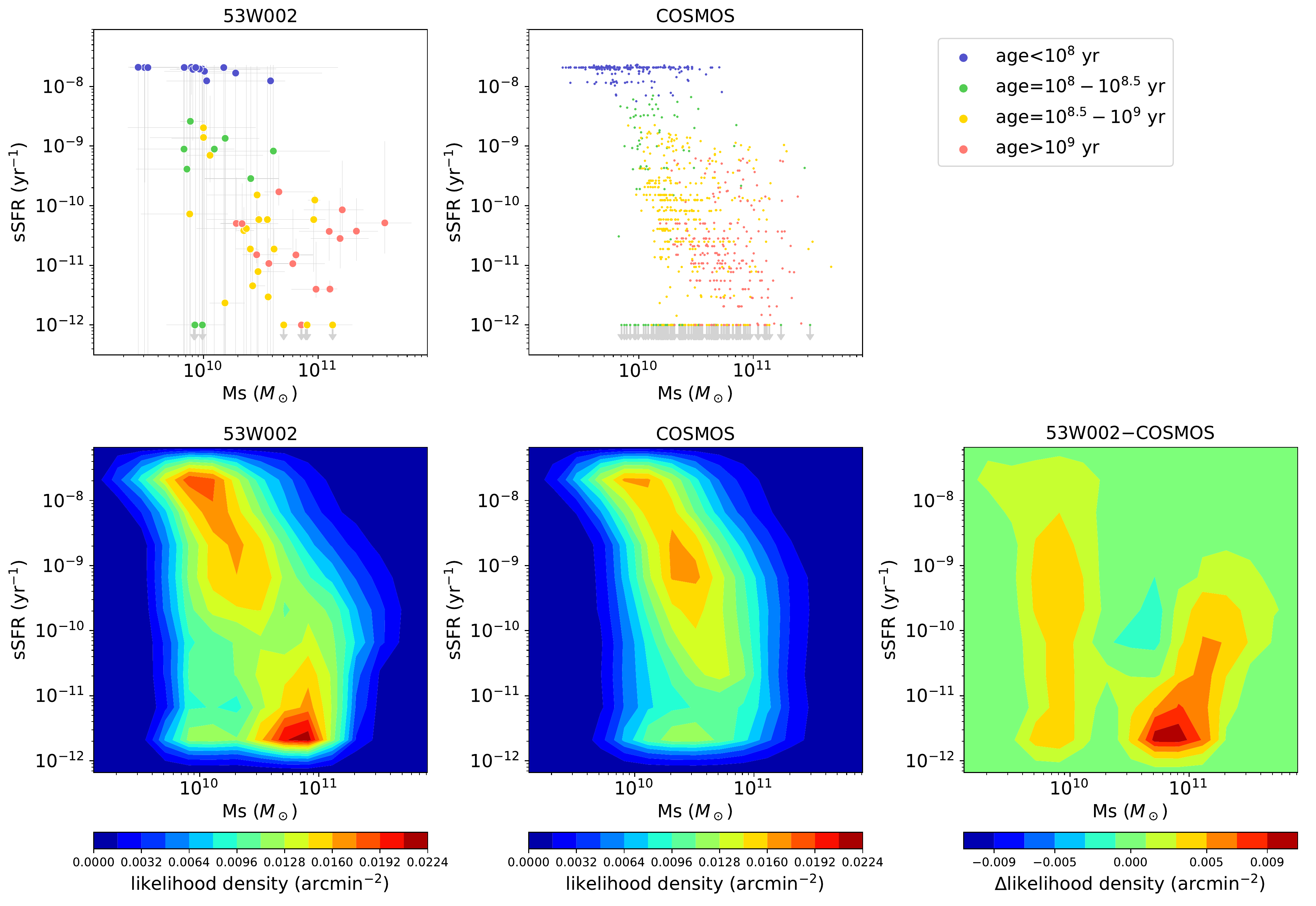}
  \caption{The same as Figure \ref{fig:fig8}, but the nebular emission is included in the SED fitting.}
  \label{fig:fig18}
\end{figure*}

\subsection{Physical Properties}\label{appendix:fitting}

In this subsection, we investigate the effects of the nebular emission in the model templates on the physical properties estimated by the SED fitting.
As in Section \ref{subsec:phys_prop}, we carried out the SED fitting fixing redshift to $z=2.39$ to estimate the physical properties of the sample galaxies by using the model templates with the nebular emission.
Model templates are the same as in Section \ref{appendix:selection}.
We used the original sample, which were selected by the SED fitting without the nebular emission, in order to compare the estimated physical properties in the case with and without the nebular emission for the same galaxies.

\begin{figure*}[ht!]
  \centering
  \includegraphics[width=17cm]{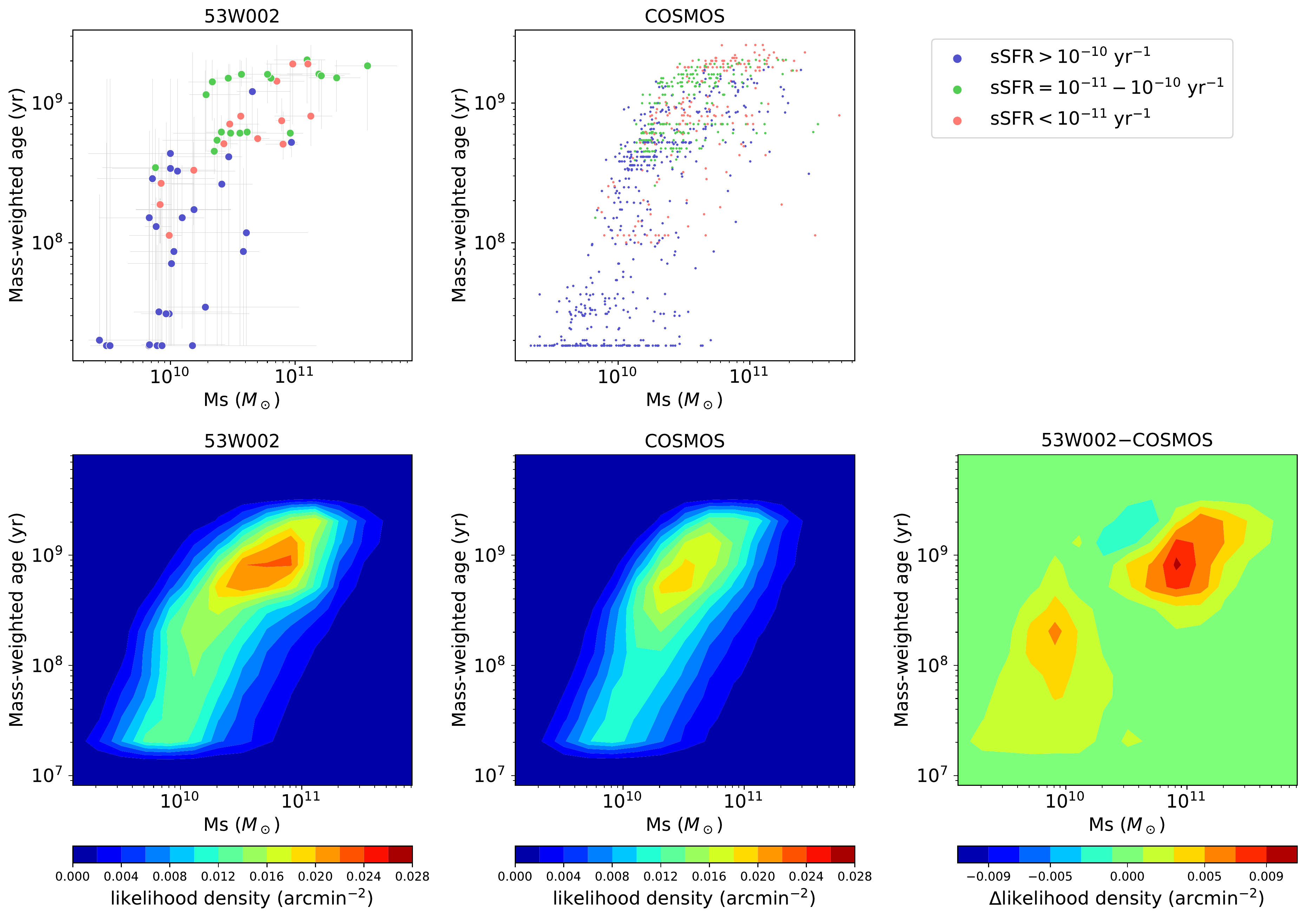}
  \caption{The same as Figure \ref{fig:fig9}, but the nebular emission is included in the SED fitting.}
  \label{fig:fig19}
\end{figure*}
Figure \ref{fig:fig17} shows the stellar mass distribution of the $JHK_s$-selected galaxies in the cases with and without the nebular emission.
The stellar mass of relatively massive galaxies with $M_s>10^{10.5}\ M_\odot$ is not significantly affected, while that of low-mass galaxies with $M_s<10^{10.5}\ M_\odot$ tend to be lower when the nebular emission is included.
Figure \ref{fig:fig18} shows the scatter plot of sSFR vs.~$M_s$, probability density map, and density excess map for the $JHK_s$-selected galaxies in the case with the nebular emission in the SED fitting.
The number and probability density of galaxies with sSFR $>10^{-8}\ \rm{yr^{-1}}$ increase compared to the case where the nebular emission is not included.
At $z=2.39$, strong emission lines of [OII]$\lambda$3727, [OIII]$\lambda\lambda4959,5007$, and H$\alpha$ enter into the $J$, $H$, and $K_s$ bands, respectively.
These emission lines could affect the broad-band photometry, especially for young star-forming galaxies, where equivalent widths of these lines tend to be large.
Since star-forming galaxies at $z\gtrsim2$ often show strong [OIII]$\lambda\lambda4959,5007$ and H$\alpha$ lines relative to the [OII]$\lambda3727$ line \citep[e.g.,][]{Shapley_2015,Steidel_2016}, these emission lines could redden $J-K_s$ (and $J-H$) color keeping $H-K_s$ color nearly constant.
Such the effects of strong emission lines could mimic that of the Balmer/4000$\rm{\AA}$ break in the photometric SED.
Indeed, the uncertainty of stellar mass and sSFR for relatively low-mass galaxies becomes larger due to this degeneracy when the nebular emission is included (upper panels of Figure \ref{fig:fig18}).
Furthermore, a significant fraction of these low-mass galaxies with $M_s\sim10^{10}\ M_\odot$ are fitted with very young models with sSFR $\sim10^{-8}\ \rm{yr^{-1}}$ in the both fields.
Since the strength of this effect seems to be similar in both fields, the probability density excess for low-mass galaxies is not significantly affected by this effect (lower right panel of Figure \ref{fig:fig18}).
On the other hand, the effect of the nebular emission does not strongly change the estimation of the physical properties for the massive galaxies probably because of their small equivalent widths of these lines.

Figure \ref{fig:fig19} is the same as Figure \ref{fig:fig18}, but for mass-weighted age vs.~$M_s$.
One can see that a significant fraction of low-mass galaxies shows young ages of $<10^{8}$ yr in the both fields.
Again, the effect is similar in the both fields, and the probability density excess for low-mass galaxies is not strongly affected.
The mass-weighted ages of massive galaxies do not significantly change when the nebular emission is included in the fitting, although there are a few galaxies with younger ages than those in the original analysis at $M_s\sim10^{10.5}\ M_\odot$.

Thus the nebular emission could affect the stellar mass, sSFR, and mass-weighted age of low-mass galaxies.
We need spectroscopic confirmation of the emission lines and/or the photometric information at longer wavelength to resolve the degeneracy.
However, the effects similarly work for those in the 53W002 and COSMOS fields, and therefore the density excesses for low-mass galaxies are not changed so strongly.
On the other hand, the estimated physical properties of massive galaxies with $M_s>10^{10.5}\ M_\odot$ are not sensitive to the nebular emission unless they have strong emission lines from AGNs.

\section{Effect of fixing redshift to $z=2.39$}\label{appendix:fixed_redshift}
In Section \ref{subsec:result_prop}, we estimated the physical properties such as $M_s$, sSFR, and mass-weighted age of the comparison sample fixing their redshifts to $z=2.39$ in order to subtract the contribution of these field galaxies from the 53W002 data.
Since their redshift distribution ranges from $z\sim1$ to $z\sim3$ (Figure \ref{fig:fig3}), fixing to erroneous redshifts for foreground/background galaxies could systematically affect the distributions of physical properties of these galaxies.
For reference, we here present the results of those galaxies with photometric redshifts from the COSMOS2015 catalog of $z_{COS}=2.39\pm0.20$.
Out of 896 galaxies in the comparison sample, 301 galaxies have $z_{COS}=2.39\pm0.20$.
In addition to the case of fixing redshift to $z=2.39$, we performed the same SED fitting analysis fixing redshift to $z_{COS}$ values.
\begin{figure*}[ht!]
  \centering
  \includegraphics[width=15cm]{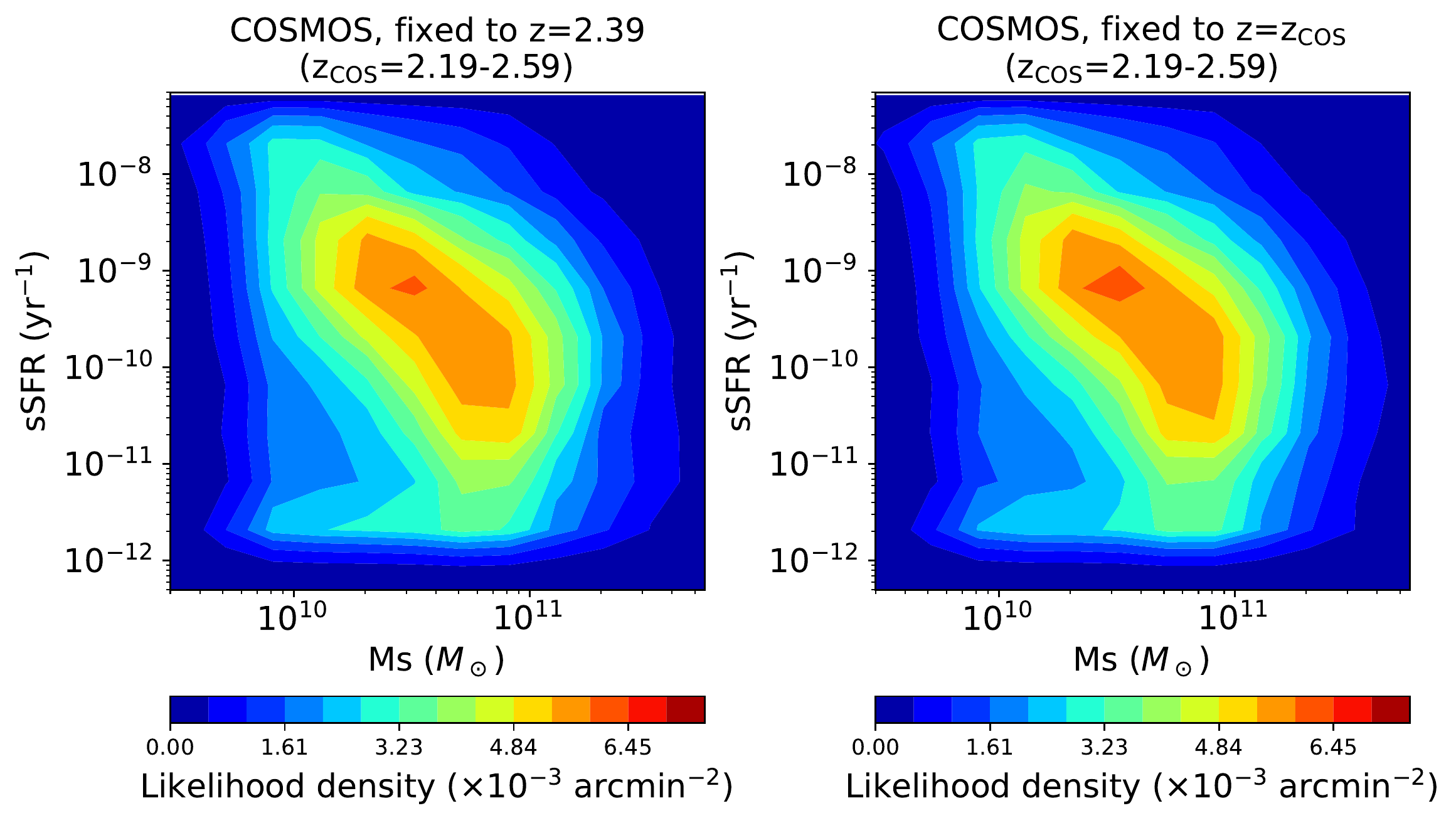}
  \caption{The probability density distributions of the $JHK_s$-selected galaxies with $z_{\rm{COS}}=2.39\pm0.20$ in the cases where their redshifts are fixed to $z=2.39$ (left panel) and $z=z_{\rm{COS}}$ (right panel) in the SED fitting.
  The color scale is corrected by multiplying that of Figure \ref{fig:fig8} for the ratio between the numbers of objects (301/896)}
  \label{fig:fig20}
\end{figure*}
\begin{figure*}[ht!]
  \centering
  \includegraphics[width=15cm]{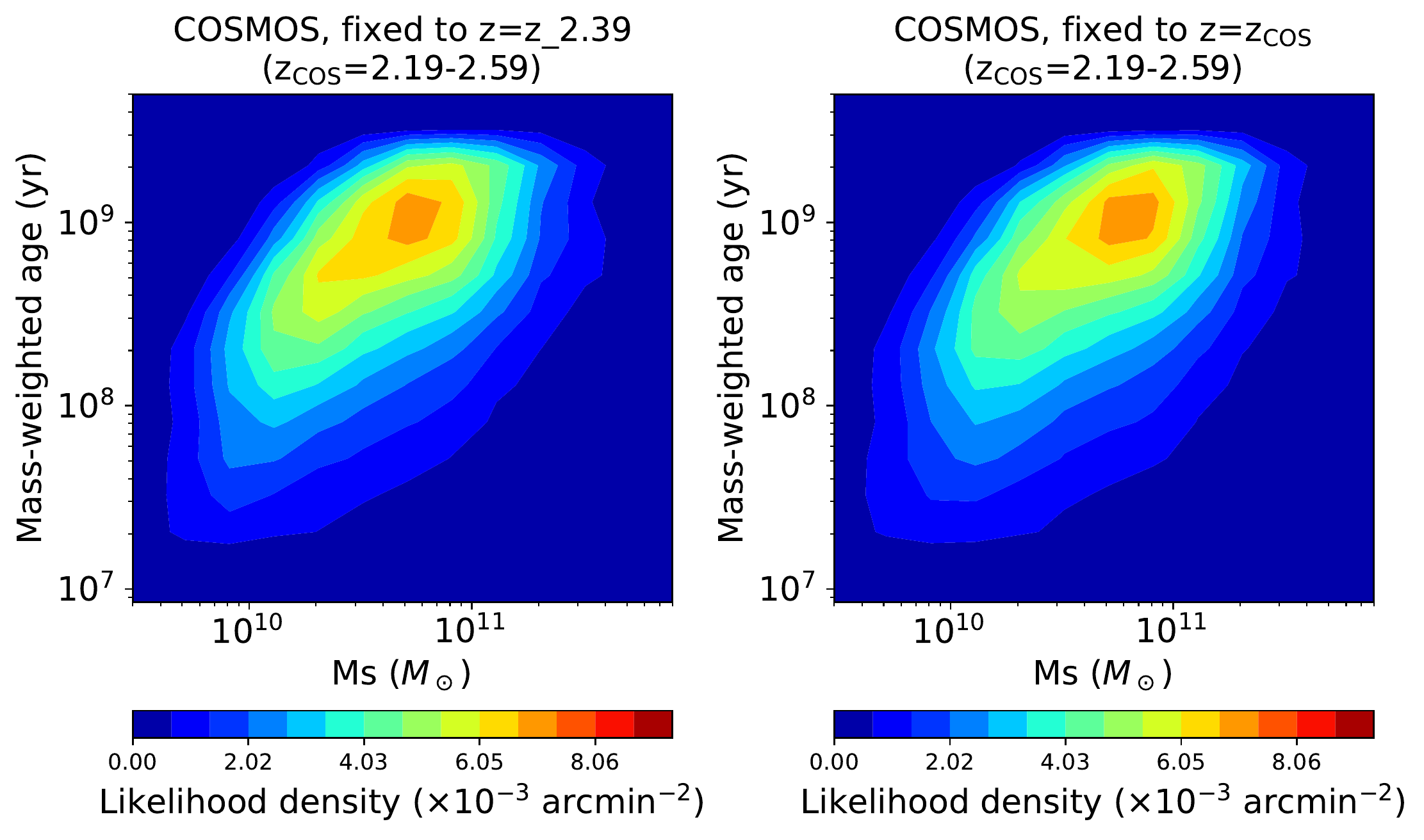}
  \caption{The same as Figure \ref{fig:fig20}, but for mass-weighted age vs.~$M_s$.}
  \label{fig:fig21}
\end{figure*}

The left and right panels of Figure \ref{fig:fig20} show the probability density distribution of sSFR vs.~$M_s$ estimated by the SED fitting fixing redshift to $z=2.39$ and $z=z_{\rm{COS}}$, respectively.
We note that the color scale is corrected by multiplying that of Figure \ref{fig:fig8} by a ratio of the numbers of objects, i.e., 301/896.
It can be seen from the left panel that the fraction of galaxies with $M_s\sim10^{10.5}$--$10^{11}\ M_\odot$, in particular, those with sSFR $=10^{-11}$--$10^{-10}\ \rm{yr^{-1}}$ increases when we limit the field galaxies shown in Figure \ref{fig:fig8} (d) to those with $z_{COS}=2.19$--$2.59$.
The probability density around $M_s\sim10^{10}\ M_\odot$ and sSFR $\sim10^{-11}$--$10^{-10}\ \rm{yr^{-1}}$ becomes lower than that in Figure \ref{fig:fig8} (d).
This is probably because the contaminants from foreground/background have relatively low stellar masses when their redshifts are fixed to $z=2.39$.
Except for these differences, the overall distributions are similar with each other.
Adopting $z_{COS}$ rather than fixing to $z=2.39$ does not change the distribution for those with $z_{COS}=2.39\pm0.20$.

The left and right panels of Figure \ref{fig:fig21} show the probability density distributions of mass-weighted age vs.~$M_s$ estimated by the SED fitting where redshift is fixed to $z=2.39$ and $z=z_{\rm{COS}}$, respectively.
The color scale is corrected for the difference in the number of objects.
The field galaxies at $z_{COS}=2.39\pm0.20$ show a higher fraction of those with $M_s\sim10^{10.5}$--$10^{11}\ M_\odot$ and age $>10^9\ \rm{yr}$ than all the comparison sample shown in Figure \ref{fig:fig9}.
The foreground/background galaxies seem to show systematically younger ages.
As in sSFR vs.~$M_s$ plane, the probability density distribution of mass-weighted age vs.~$M_s$ does not significantly change even if we adopt the photometric redshifts from COSMOS2015 catalog.

\section{Detection Completeness}\label{appendix:comple}
In Section \ref{subsec:completeness}, we carried out the simulations to estimate the detection completeness for $JHK_s$-selected galaxies on the $K_s$-band image, which is used to correct the number density of the comparison sample.
We here describe details of the procedures in the simulations.

\begin{figure*}[ht!]
  \centering
  \includegraphics[width=17cm]{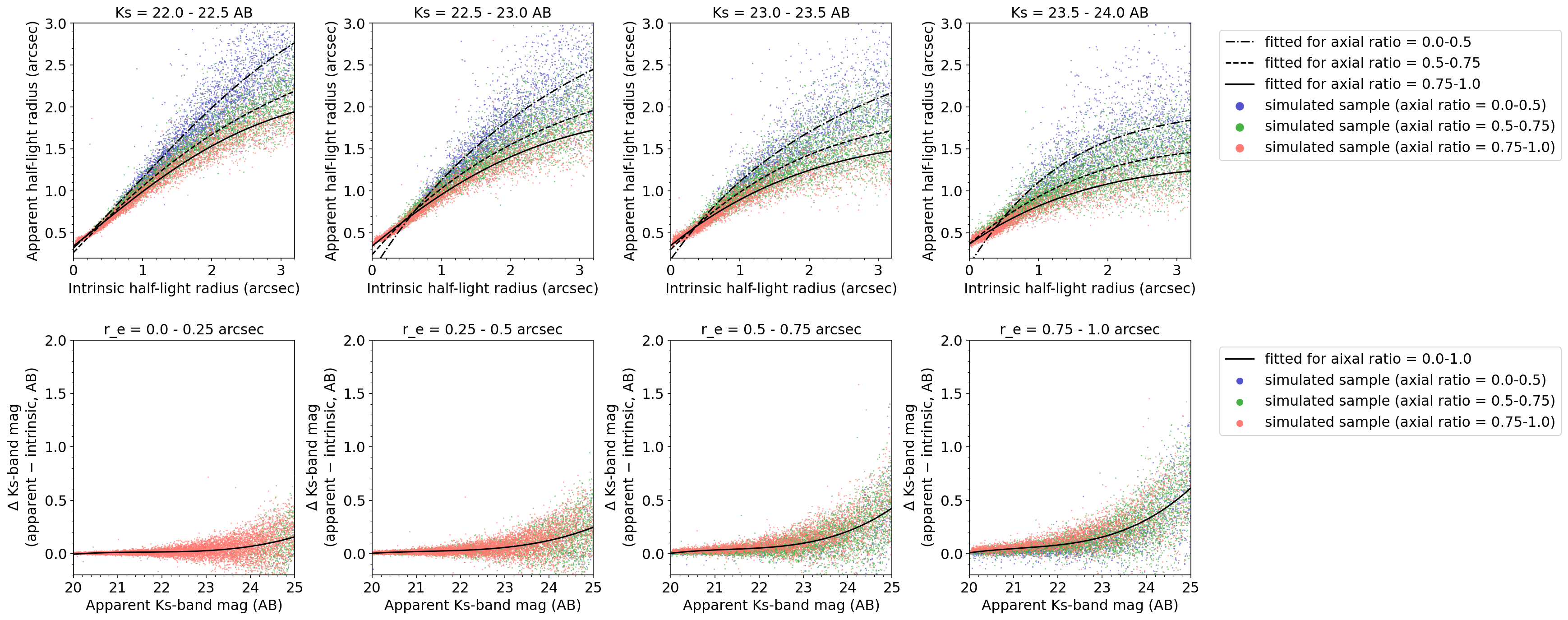}
  \caption{Upper panels: Apparent vs.~intrinsic half-light radii for the artificial objects with different $K_s$-band magnitudes and axial ratios. 
  Blue, green, and red dots represent the artificial objects with the axial ratio of 0.0--0.5, 0.5--0.75, and 0.75--1.0, respectively. 
  Dash-dotted, dashed, and solid lines represent the fitting results with 4th-order polynomial between the apparent and intrinsic half-light radii for the objects with an axial ratio of 0.0--0.5, 0.5--0.75, and 0.75--1.0, respectively.
  Lower panels: Difference between the apparent and intrinsic $K_s$-band magnitudes vs.~$K_s$-band magnitude for the artificial objects with the different half-light radii.
  Solid lines show the fitting results with 3rd-order polynomial for the objects with axial ratio of 0.0--1.0.}
  \label{fig:fig22}
\end{figure*}

\begin{figure*}[ht!]
  \centering
  \includegraphics[width=15cm]{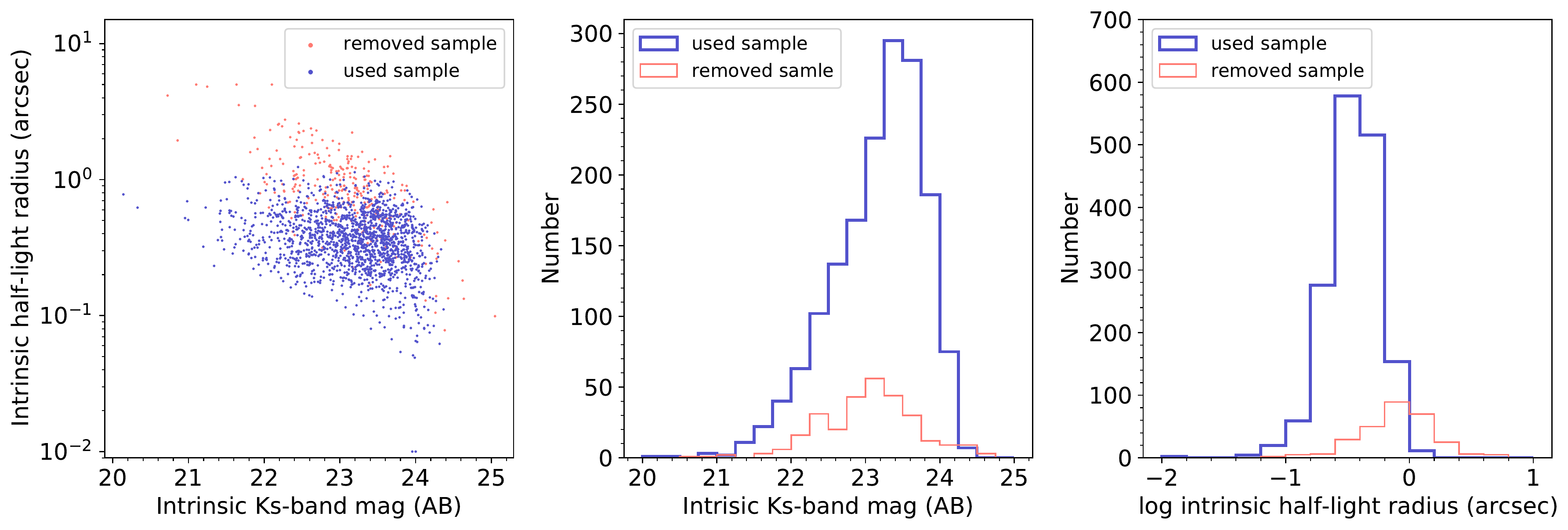}
  \caption{Left: Intrinsic half-light radius vs.~$K_s$-band magnitude for the $JHK_s$-selected galaxies in the COSMOS field. 
  Blue and red dots indicate the $JHK_s$-selected galaxies used and removed due to the nearby bright sources in the simulations, respectively.
  Middle: The distributions of the intrinsic $K_s$-band magnitude for the objects. The colors are the same as the left panel.
  Right: The same as the middle panel, but for the intrinsic half-light radius.}
  \label{fig:fig23}
\end{figure*}

At first, we need to estimate the intrinsic sizes of $JHK_s$-selected galaxies detected in the COSMOS field as a function of $K_s$-band magnitude in order to make artificial objects with similar surface brightness with those galaxies in the main simulations.
In this analysis, we included the $JHK_s$-selected galaxies with $K_s>23.3$ to examine the completeness around the magnitude limit.
We ran SExtractor on the $K_s$-band image of the UltraVISTA DR2 and measured apparent half-light radii of $JHK_s$-selected galaxies.
We excluded those galaxies significantly affected by the nearby bright sources from the analysis.
We then carried out the simulations with {\tt{IRAF/MKOBJECTS}} on the same $K_s$-band image to derive the relation between apparent and intrinsic sizes of objects.
In the simulations, artificial objects with various magnitudes ($K_s=20-25$), intrinsic half-light radii ($r_e=0\farcs0$--$5\farcs0$), and axial ratio (0.1--1.0) were convolved with PSF of the image and added to random positions in the $K_s$-band image.
We assumed exponential surface brightness profiles for these artificial objects.
We ran SExtractor to measure the apparent half-light radii of the added artificial objects.
We repeated these procedures and carried out 250,000 such simulations in total.
The upper panels of Figure \ref{fig:fig22} show comparison between the apparent and intrinsic half-light radii for artificial objects with different $K_s$-band magnitudes and axial ratios.
Note that both the apparent and intrinsic half-light radii are scaled to the semi-major axis in the figure.
We fitted the relation between the apparent and intrinsic half-light radii for objects in each magnitude and axial ratio range with 4th-order polynomial (curves in the upper panels).
We used the fitted functions for the magnitude and axial ratio bins to convert the apparent half-light radii of the $JHK_s$-selected galaxies in the COSMOS field to the intrinsic half-light radii.
We also investigated the relation between the measured (MAG\_AUTO from SExtractor) and intrinsic $K_s$-band magnitudes by using the same simulations with the artificial objects (the lower panels of Figure \ref{fig:fig22}).
Although differences between the measured and intrinsic magnitudes are small at $K_s<23.3$, we similarly fitted the relation for artificial objects in each half-light radius range with 3rd-order polynomial, and converted the measured magnitudes of the $JHK_s$-selected galaxies to the intrinsic $K_s$-band magnitudes.
Figure \ref{fig:fig23} shows the distribution of the estimated intrinsic half-light radii and $K_s$-band magnitudes of the $JHK_s$-selected galaxies in the COSMOS field.

Next, we added artificial objects with the similar intrinsic half-light radii as the $JHK_s$-selected galaxies in the COSMOS field to the $K_s$-band image of the 53W002 field.
For each artificial object with a given (intrinsic) $K_s$-band magnitude, we randomly selected one $JHK_s$-selected galaxy in the COSMOS field in the same magnitude range, and adopted its intrinsic half-light radius.
We chose to fix the intrinsic axial ratio to 0.8, because the resultant distribution of the measured axial ratio of the detected artificial objects in this setup reproduced well the observed distribution of the axial ratio of $JHK_s$-selected galaxies in the 53W002 field (lower panels of Figure \ref{fig:fig25}).
We used {\tt{IRAF/MKOBJECTS}} to convolve such artificial objects with the PSF of the $K_s$-band image of the 53W002 field and add them to random positions in the sky image described in Section \ref{subsec:detection}.
We then ran SExtractor with the same surface brightness threshold and other parameters as in Section \ref{subsec:detection}, and examined whether the artificial object was detected or not with the SEGMENTATION image from SExtractor.
We carried out 500 such simulations and calculated the fraction of the detected artificial objects for each magnitude bin.
Since the PSF and depth varies among the different fields of view and chips, we did the procedures mentioned above for each field of view and chip separately.

\begin{figure}[ht!]
  \centering
  \includegraphics[width=15cm]{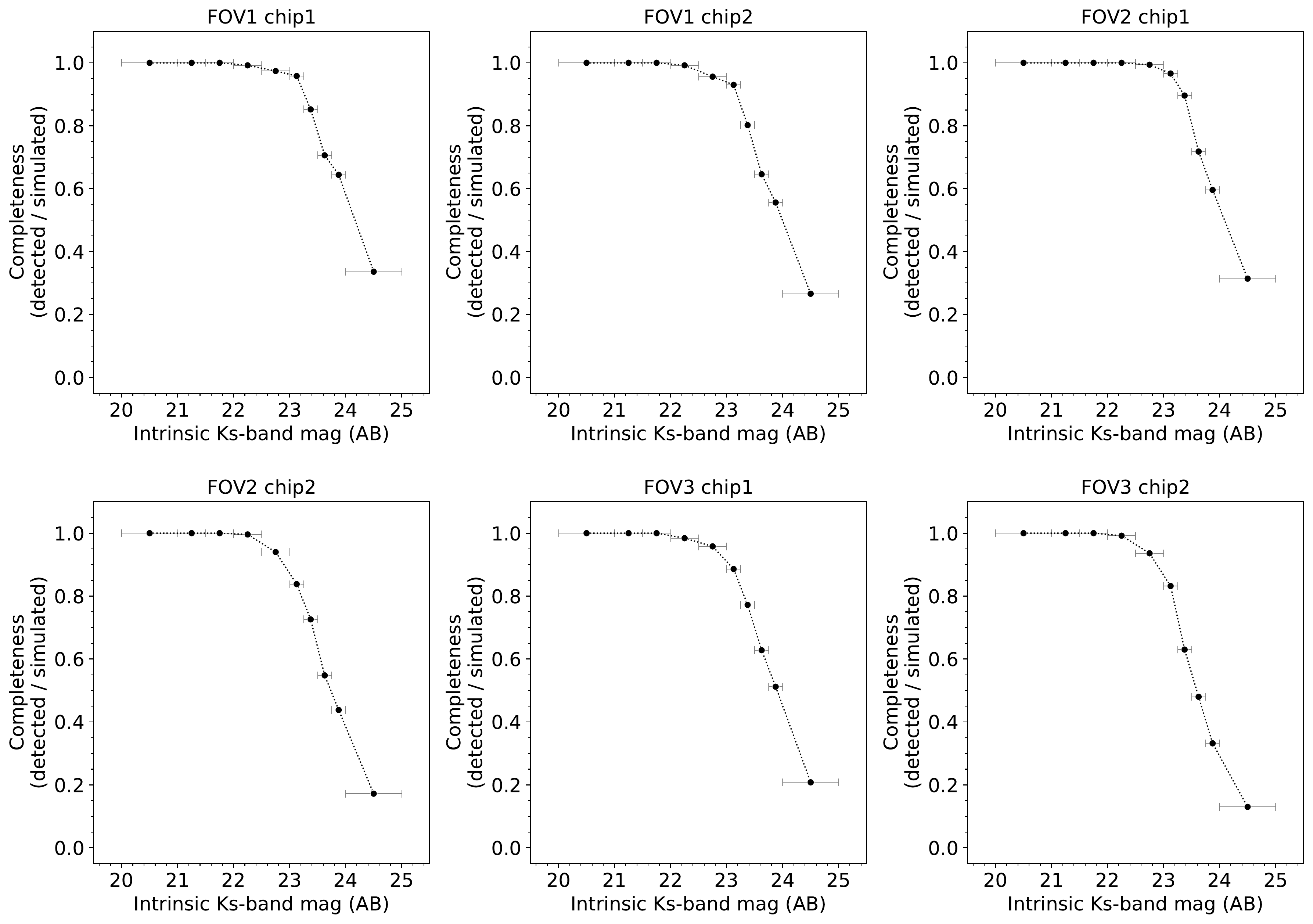}
  \caption{Detection completeness as a function of intrinsic $K_s$-band magnitude in each field of view and chip in the 53W002 field. 
  The error bars represent widths of the magnitude bins.}
  \label{fig:fig24}
\end{figure}

\begin{figure}[ht!]
  %\centering
  \includegraphics[width=18cm]{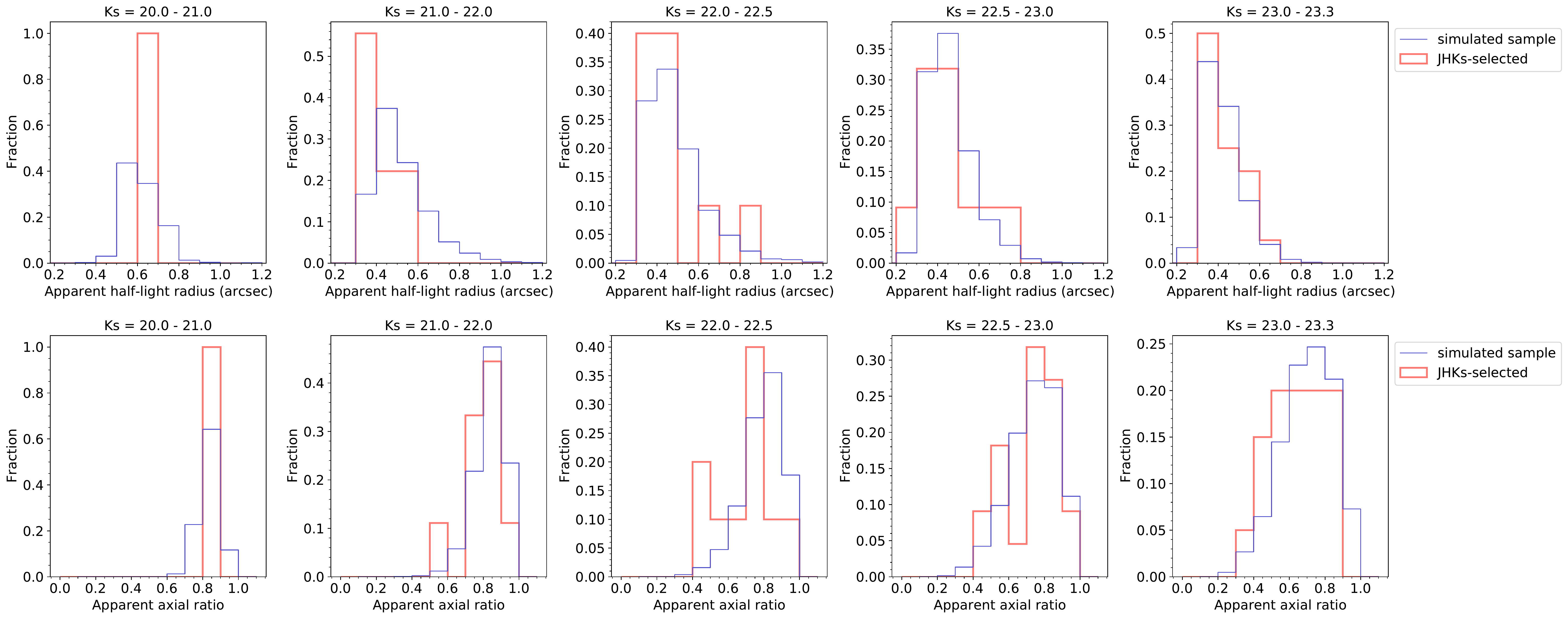}
  \caption{Upper panels: Distributions of apparent half-light radius for the artificial objects in the simulations (blue) and the observed $JHK_s$-selected galaxies in the 53W002 field (red).
  Lower panels: The same as the upper panels, but for apparent axial ratio.
  }
  \label{fig:fig25}
\end{figure}

Figure \ref{fig:fig24} shows the estimated detection completeness for all the fields of view and chips in the 53W002 field.
While the completeness is slightly higher in FOV1 observed in the better seeing condition, the differences among the fields of view and chips are relatively small.
We calculated a weighted average of the results in all the fields of view and chips using the effective area of each field of view and chip as a weighting factor, and adopted it as the detection completeness in the 53W002 field (Figure \ref{fig:fig4}).

Finally, we checked the apparent sizes and  axial ratios of the detected artificial objects in the simulations as a function of $K_s$-band magnitude and compared them with those of the observed $JHK_s$-selected galaxies in the 53W002 field (Figure \ref{fig:fig25}).
There is no large systematic difference in the distributions, although the number of objects is small at bright magnitudes.

\section{Stellar mass completeness limit}\label{appendix:Mslim}
\begin{figure}[ht!]
  \centering
  \includegraphics[width=17cm]{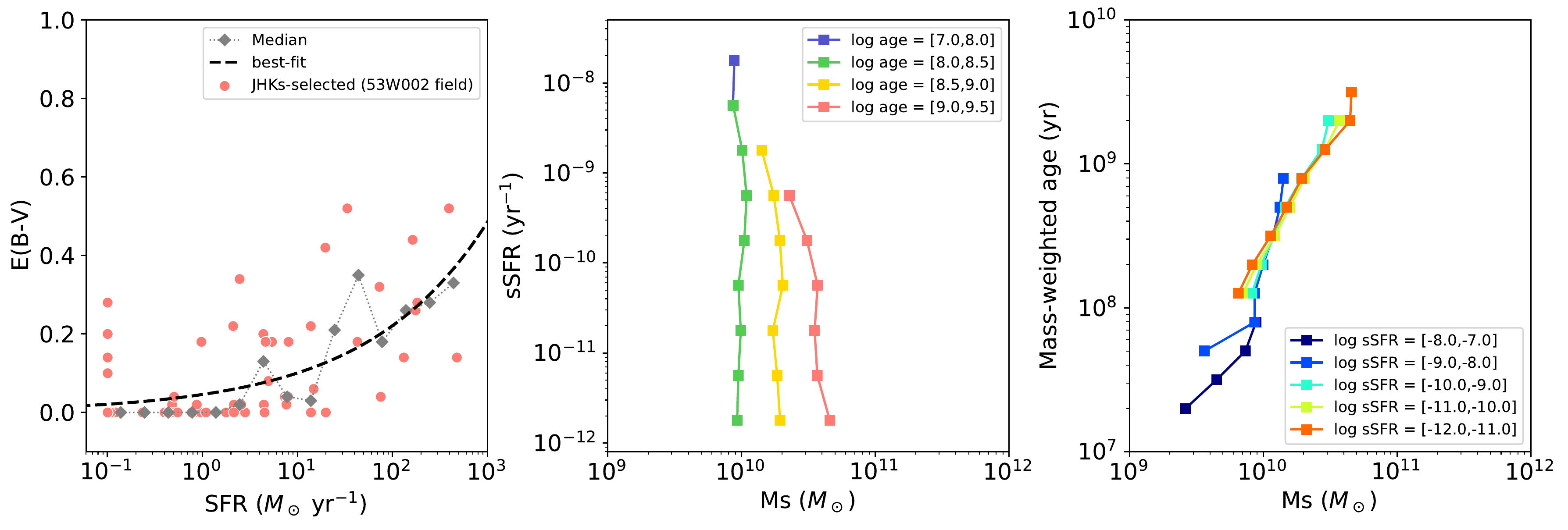}
  \caption{Left: Relation between $E(B-V)$ and SFR for the $JHK_s$-selected galaxies in the 53W002 field. 
  Red circles show the $JHK_s$-selected galaxies. 
  Grey squares and dotted line indicate median values of $E(B-V)$ in SFR bins with a width of 0.25 dex. 
  Black dashed line shows the fitting result for the median values with a form of $E(B-V)=a\times{\rm{SFR}}^{b}$.
  Middle: Stellar mass completeness limits on sSFR vs.~$M_s$ plane.
  Symbols and lines with the different colors show those for the different mass-weighted ages.
  Right: The same as the middle panel, but for the different sSFRs on mass-weighted age vs.~$M_s$ plane.
  }
  \label{fig:fig26}
\end{figure}

We here describe a method to estimate the stellar mass completeness limit shown in Figures \ref{fig:fig7}--\ref{fig:fig9}.
We define the completeness limit as a stellar mass above which all galaxies at $z=2.39$ are brighter than the magnitude limit of $K_s=23.3$.
We calculated the stellar masses of the model templates with $K_s=23.3$ and $z=2.39$, and adopted a maximum value as the completeness limit.
We used the same model templates from the GALAXEV library as in the SED fitting (Section \ref{subsec:sample}).
Since stellar mass to luminosity ratio depends on distribution of stellar age in the galaxy, we calculated the completeness limit as a function of mass-weighted age and sSFR.
We assumed an $E(B-V)$ value of each model template estimated from empirical relation between $E(B-V)$ and SFR for the $JHK_s$-selected galaxies shown in the left panel of Figure \ref{fig:fig26}.
We fitted median values of $E(B-V)$ for $JHK_s$-selected galaxies in the 53W002 field as a function of SFR with a function of $E(B-V)=a\times{\rm{SFR}}^{b}$.
Since the $JHK_s$-selected galaxies in the COSMOS field show the almost same relation between $E(B-V)$ and SFR, we used the relation in the left panel of Figure \ref{fig:fig26} for the both fields.
We calculated the maximum stellar mass in the model templates with $K_s=23.3$ and $z=2.39$ for each mass-weighted age and sSFR range. 

The middle and right panels of Figure \ref{fig:fig26} show the estimated stellar mass completeness limit on the sSFR vs.~$M_s$ and mass-weighted age vs.~$M_s$ planes, respectively.
From these panels, one can see that the completeness limit mainly depends on the mass-weighted age, and its dependence on sSFR is relatively weak for a given mass-weighted age.
Therefore, we also estimated the completeness limit as a function of mass-weighted age regardless of sSFRs and used it in Figure \ref{fig:fig7} and \ref{fig:fig9}, while we used those as a function of sSFR for the different mass-weighted ages shown in the middle panel of Figure \ref{fig:fig26} for Figure \ref{fig:fig8}.
While those galaxies with stellar mass larger than the completeness limit should be basically brighter than $K_s=23.3$, we note that the scatter around the assumed relation between $E(B-V)$ and SFR could affect the completeness around the limit to some extent.

%% the bibliography. The sample631.bib file was populated from ADS. In order to

%\bibliography{sample631}{}
%\bibliographystyle{aasjournal}

%\clearpage



\end{document}